\documentclass[authoryear,preprint,review,sort&compress]{cas-sc}
\usepackage[utf8]{inputenc}
\usepackage[USenglish]{babel}
\usepackage{amsmath}
\usepackage{mathtools}
\usepackage{geometry}
\usepackage{graphicx}
\usepackage{amssymb}
\usepackage{natbib}
\usepackage{fleqn}
\usepackage{subcaption}
\usepackage{multirow}
\usepackage{framed}
\usepackage{setspace}
\graphicspath{ {./Figures/} }
\usepackage{graphicx, color}

\begin{document}

\shorttitle{}
\shortauthors{van Esch et~al.}

\title[mode=title]{A Bayesian method for inference of effective connectivity in brain networks for detecting the Mozart effect \footnote{\copyright <2020>. This manuscript version is made available under the CC-BY-NC-ND 4.0 \\
\indent license http://creativecommons.org/licenses/by-nc-nd/4.0/} } 

\author[1]{Rik J. C. van Esch}

\author[1]{Shengling Shi }
\cormark[1]
\ead{s.shi@tue.nl}

\author[1]{Antoine Bernas}

\author[1]{Svitlana Zinger}

\author[1,2,3]{Albert P. Aldenkamp}

\author[1]{Paul M. J. Van den Hof}

\address[1]{Department of Electrical Engineering, Eindhoven University of Technology, Groene Loper 19, 5612 AP Eindhoven, The Netherlands}

\address[2]{Department of Neurology, Maastricht University Medical Center, Universiteitssingel 40, 6229 ER Maastricht, The Netherlands}

\address[3]{Department of Behavioral Sciences, Epilepsy Center Kempenhaeghe, Sterkselseweg 65, 5591 VE Heeze, The Netherlands}

\cortext[cor1]{Corresponding author}
\date{\today}

\begin{abstract}[S U M M A R Y]
\doublespacing
Several studies claim that listening to Mozart music affects cognition and can be used to treat neurological conditions like epilepsy. Research into this Mozart effect has not addressed how dynamic interactions between brain networks, i.e. effective connectivity, is affected. Granger-causality analysis is often used to infer effective connectivity. First, we investigate if a new method, Bayesian topology identification, can be used as an alternative. Both methods are evaluated on simulation data, where the Bayesian method outperforms the Granger-causality analysis in the inference of connectivity graphs of dynamic networks, especially for short data lengths. In the second part, the Bayesian method is extended to enable the inference of changes in effective connectivity between groups of subjects. Next, we apply both methods to fMRI scans of 16 healthy subjects, who were scanned before and after exposure to Mozart’s sonata K448 at least 2 hours a day for 7 days. Here, we investigate if the effective connectivity of the subjects significantly changed after listening to Mozart music. The Bayesian method detected changes in effective connectivity between networks related to cognitive processing and control: First, in the connection from the central executive to the superior sensori-motor network. Second, in the connection from the posterior default mode to the fronto-parietal right network. Finally, in the connection from the anterior default mode to the dorsal attention network, but only in a subgroup of subjects with a longer listening duration. Only in this last connection an effect was found by the Granger-causality analysis.
\end{abstract}
\begin{keywords}
    fMRI \sep Neurodynamics \sep Resting-state networks \sep Bayesian model selection \sep Mozart effect \sep ICA
\end{keywords}
\maketitle 
\doublespacing
\section{Introduction}
In the last decade the effects of music on the brain have been an active topic of research \citep{MusicFMRI,MusicWhitehead,CbfMusic}. In particular one piece of classical music, Mozart's Sonata K448, has been claimed to have unique effects on the brain, besides the general effects caused by music. One of the original studies by \citet{MozartRauscher} reports that listening to Mozart music increases spatial and temporal reasoning skills in healthy subjects, but this effect did not last beyond the 15 minute testing period. However, a strong effect specific to Mozart music was unable to be replicated in repeated studies \citep{MozartMeta}. In another study \citep{MozartEEG} neural activity was measured through spectral analysis of electroencephalogram (EEG) data. After listening to Mozart music, an increase in brain wave activity linked to memory, cognition and problem solving was observed in adults.\par
Besides these effects, Mozart music has also been reported as being effective as a medicine. First of all, \citet{MozartCoppola} reports that consistent exposure to Mozart music for 15 days caused a reduction in the frequency of epileptic seizures in children with epilepsy and in some cases led to complete recession of seizures. Furthermore, the effects persisted even after the exposure to Mozart music was stopped. Another study \citep{MozartBodner} reported similar positive results in adults with epilepsy. Finally, in a study of the influence of Mozart music therapy on schizophrenia patients \citep{MozartYang}, Mozart music was found to have effects on the functional connectivity of patients, which was inferred from functional magnetic resonance imaging (fMRI) data.\par
However, no well-established theories exist on the neural dynamics in the brain that are responsible for the Mozart effect found in theses studies. Moreover, research into the existence of a Mozart effect does not focus on inferring effective connectivity between brain networks from time series, so in this work we will focus on methods that can infer how brain networks are connected and interact, using fMRI BOLD time series that describe the dynamic neural activity from those brain networks.\par
The brain is a dynamic system; cognition and consciousness arises from not just the static correlations between brain networks, i.e. functional connectivity, but also through the dynamic behavior of the brain. Therefore, the inference of  dynamic interactions of brain networks, i.e. effective connectivity, is an important aspect in the analysis of the brain as it reveals how brain networks dynamically influence each other. 
In this work we focus on the inference of effective connectivity using resting-state fMRI BOLD measurements. We could infer connectivity directly from voxel BOLD measurements \citep{FmriVoxel} or from the mean BOLD signals in brain networks defined by an atlas \citep{fMRIAtlas}. Instead, we will follow an approach that uses independent component analysis (ICA) \citep{gICA,ICAhyvar} to find active brain networks and their corresponding ICA time series. These ICA time series are then used as an indication of dynamic behaviour of that brain region \citep{GCBernas,GrangerICAlondei,GrangerICAdemirci,GrangerICAregner}.\par
\subsection{Literature review of related estimation methods}
By far the most popular method of inference of effective connectivity is the Granger-causality analysis \citep{originGranger,GrangerGeweke}. While the method originated in the field of economics, it has been used in the past for the inference of brain connectivity using EEG \citep{GCEEG} and fMRI data of the brain \citep{GCBernas}. Its strength is the potential as an exploratory method to assess the effective connectivity between a large number of brain networks of interest, as effective connectivity can be assessed pairwise between brain networks. Despite the popularity of this method, it is not without flaws. Most importantly, care must be taken when performing the Granger-causality analysis on fMRI data, as the low sampling rate of fMRI measurements can make the method unreliable \citep{GCdowns}.\par
Next, DCM \citep{DCMmain} is a modelling approach, which 
%The first approach, DCM \citep{DCMmain} 
is specifically designed to model neural activity \citep{Balloonwindkessel} and effective connectivity using specialized state-space models, for example in \citep{DCMtwostate} excitatory and inhibitory states are used, which is in line with how populations of neurons function in the brain. DCM allows for much more biologically accurate modelling of the neural activity compared to Granger-causality analysis, but the estimation of an accurate DCM model can be computationally expensive and is often limited to modelling the interactions between a small number of brain networks. In addition, DCM typically requires prior knowledge regarding the network connectivity, while this prior knowledge may not always be available \citep{vicente2011transfer}. For this reason, DCM is not often used as an exploratory method to assess effective connectivity and is thus also problematic for inferring the existence of a Mozart effect, as we do not have the prior knowledge about how and where the effective connectivity might be affected by listening to Mozart music.\par
A different type of methods that uses a regression model \citep{FmriVoxel,atomnorm} with a re-weighted regularization to force a subset of model parameters of a connection to zero and thus to infer a sparse estimate of the effective connectivity. These methods evaluate the effective connectivity of the complete network of brain networks, instead of the pairwise approach of the Granger-causality analysis. Although these methods scale very well with large dynamic networks, i.e. with many nodes, they often have many tuning parameters which must be carefully chosen to achieve a good estimate of the effective connectivity.\par
Furthermore, there exist non-parametric approaches, such as the method detailed by \citet{wienerMaterassi}. This method evaluates the effective connectivity pairwise between brain networks. It calculates a Wiener filter estimate of the dynamics of one connection to infer if the connection exists in the dynamic network or not. The wiener filter requires a lot of data to compute accurate estimates of the effective connectivity. The Mozart study uses fMRI measurements and the data length is relatively short. Thus, it is unlikely that we will get good Wiener filter estimates given this short data length.\par
Finally, in this work we will consider a relatively new method, Bayesian topology identification \citep{BayesShi,BayesChiuso}, which employs a Bayesian model selection approach to estimate the connectivity of a dynamic network. The main advantage of this approach is the ability to incorporate a prior probability distribution of the model, and thus the Bayesian method performs well even when not a lot of data is available. This is very useful in combination with fMRI data and the ICA procedure, where the data length is short. The Bayesian method uses a non-parametric model, and unlike the wiener filter, it determines effective connectivity of the complete network. By making use of Gaussian processes, the model order can be increased beyond what is possible with parametrized models, like those used by the Granger-causality analysis, without the risk of overfitting. This leads to more accurate estimation of the effective connectivity. 
\par
\subsection{Paper structure}
Our first research question is to investigate whether Bayesian topology identification can be used as an alternative to Granger-causality analysis for the inference of the effective connectivity using fMRI data. First, we investigate if there is a significant difference in the performances of the two methods in a simulation setup. Then, because the Bayesian method was originally developed to infer graph estimates of the brain network connectivity of one individual, we extend the Bayesian method such that we can draw conclusions on the connectivity of groups of data sets, which is beneficial in the inference of the Mozart effect.\par
After we have extended the Bayesian approach, our second research question is to investigate the influence of listening to Mozart’s Sonata K448 on effective connectivity through fMRI data of healthy adults. Both the Granger-causality analysis and the Bayesian approach are applied in the Mozart effect study, to consolidate the findings. Here, we put more emphasis on the results obtained from the approach which performs better in the simulation study. Furthermore, we will investigate if the length of time that subjects listened to Mozart music leads to different conclusions on the effective connectivity.\par
This paper is structured as follows: first, we will detail the model we will use to describe the dynamic interactions between brain regions. Second, we describe the existing Granger-causality analysis and how we can infer connectivity using fMRI data. Third, we detail the Bayesian topology identification method and develop an extension to the Bayesian method that can test the hypothesis of effective connectivity change between two groups of subjects. Then, we test the performance of the Granger-causality analysis and the Bayesian method using simulated data. Finally, we will apply the two methods to the real fMRI data and infer if listening to Mozart's Sonata K448 has an effect on the effective connectivity of the subjects in the study.
\section{Theory}
To infer brain network connectivity using fMRI measurements of the neurodynamics of the brain, we will first define a modelling framework wherein the dynamic interactions between brain networks can be modelled appropriately. 
\subsection{Dynamic network model of brain network connectivity}
The effective connectivity between brain networks, using fMRI measurements to describe neural activity, can be described by linear  models \citep{GCdowns}. Furthermore, in general the causal relationship between two brain networks can be described by an infinite impulse response (IIR) filter, as long as the hemodynamics \citep{BOLDogawa} can be considered as an invertible filter \citep{GCdowns}. The neural activity of brain networks is described by $L$ time series $w_j(t)$, $ j = 1,\ldots,L$, i.e. one ICA time series for each brain network found by an ICA decomposition of fMRI measurements of a subject. Now, the dynamic interactions between brain networks can be modeled as follows \citep{VdHofdynnet}:
\begin{equation} \label{eq:IIR}
\begin{aligned}
    & w_j(t)=\sum_{i\in I}G_{ji}(q)w_i(t)+\eta_j(t), \quad j\in I,\\
    & G_{ji}(q)=\sum_{k=1}^{\infty}\theta_{ji,k}q^{-k},
\end{aligned}
\end{equation}
where $q$ is the shift operator, i.e. $q^{-1}w_j(t)=w_j(t-1)$, $I=\{1,\ldots,L\}$, $\theta_{ji,k}$ indicates one coefficient of $G_{ji}(q)$ and $\eta_j (t)$ denotes the background noise of brain network $j$. The full model is written in matrix form as:
\begin{equation} \label{eq:IIRfull}
    w(t)=G(q)w(t)+\eta(t),
\end{equation}
where $G(q)$ is a $L\times L$ matrix, where entry $(j,i)$ is $G_{ji}(q)$, $w(t)=[w_1(t),\ldots,w_L(t)]^T$ and $\eta(t)=[\eta_1(t),\ldots,\ \eta_L(t)]^T$. \par
Some assumptions are made on the components of the model in \eqref{eq:IIR}:
\begin{enumerate}[\textbullet]
    \item $G_{ji}(q)$ is stable for all $j$,$i$, i.e. $\sum_{k=1}^\infty  | \theta_{ji,k}| < \infty$;
    \item $(I-G(q))^{-1}$ is stable, i.e. all its entries are stable;
    \item $\eta_j(t)$ is Gaussian distributed with zero mean and unknown variance $\sigma_j^2$, and is independent over $j$ and $t$.
\end{enumerate} \par
Each IIR transfer operator $G_{ji}(q)$ can be approximated by a finite-order impulse response of order $m$, which will have no significant effect on the performance of the model if $m$ is chosen large enough. The approximation is written as an auto-regressive (AR) model of order $m$:
\begin{equation}
    w_j(t)=\sum_{i=1}^L\sum_{k=1}^m w_i(t-k)\theta_{ji,k}+\eta_j(t),
\end{equation}
and written in matrix form:
\begin{equation} \label{eq:ARvecform}
    w_j=\sum_{i\in I} A_{ji}\theta_{ji}+\eta_j,
\end{equation}
where
\begin{equation*}
\begin{gathered}
    A_{ji}=
    \begin{bmatrix}
        w_i(-1)  & w_i(-2)  & \ldots & w_i(-m)   \\
        w_i(0)   & w_i(-1)  & \ldots & w_i(-m+1) \\
        \vdots   & \ddots   & \ddots & \vdots    \\
        w_i(N-1) & w_i(N-2) & \ldots & w_i(N-m)  \\
    \end{bmatrix}.
\end{gathered}
\end{equation*}
Here $w_j\in \mathbb{R}^N$, $\eta_j\in \mathbb{R}^N$ are vectors of time series of brain network $j$, $\theta_{ji}$ is the parameter vector of the connection from $w_i(t)$ to $w_j(t)$, and $A_{ji}$ is the matrix of time shifted columns of $w_i$, where $w_i(t)=0$ for $t<0$. This VAR model will be used in the inference of Granger-causality and furthermore forms the modelling framework on which the Bayesian topology identification method relies.\par
\subsection{Directed graph}
\begin{figure}
    \centering
    \includegraphics[width=9cm]{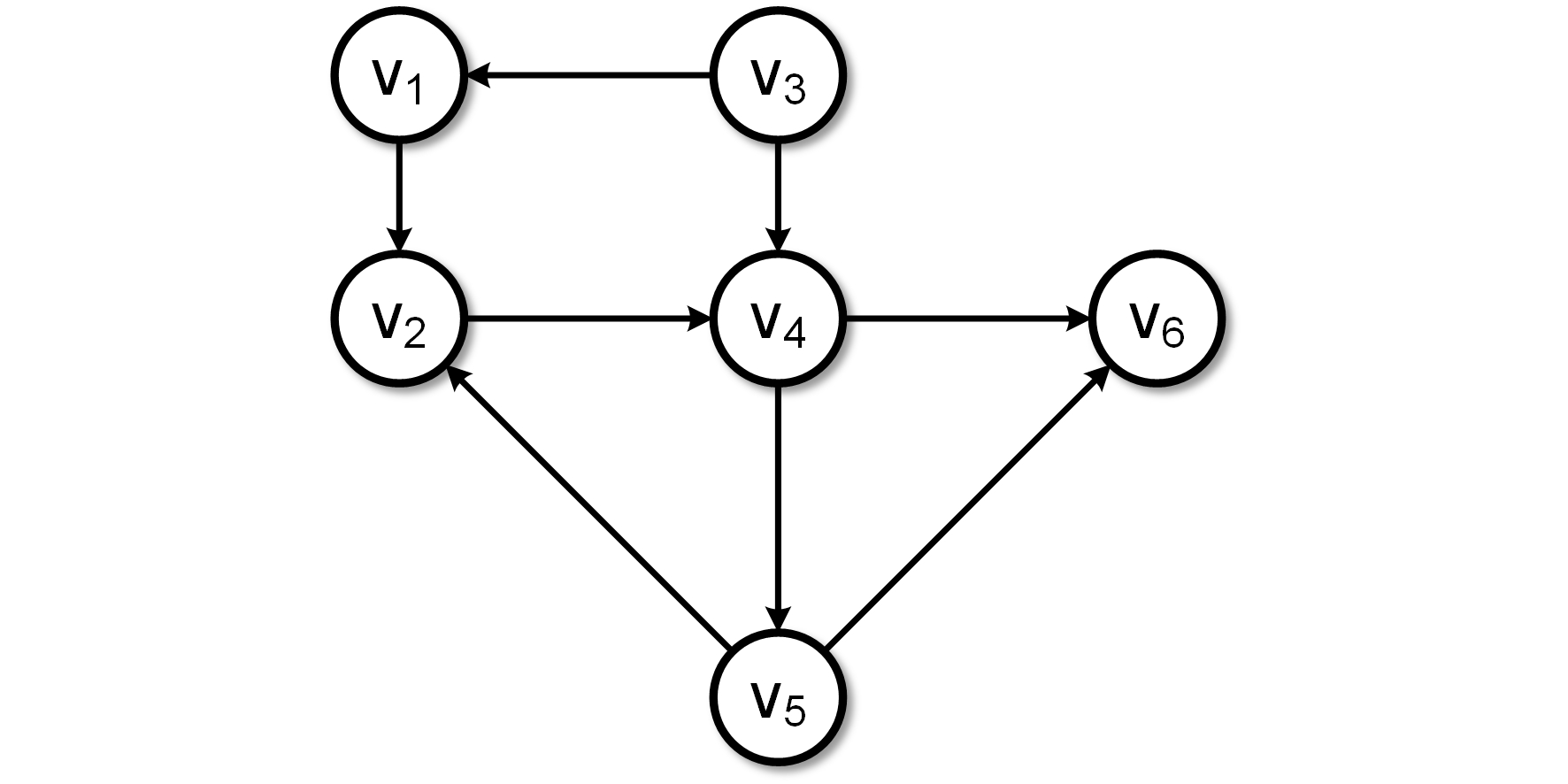}
    \caption{Example of a six-node directed graph. (This is a single column figure)}
    \label{fig:dirgraph}
\end{figure}

We can also define a graphic representation of model \eqref{eq:IIR} that encodes the effective connectivity between brain networks. A directed graph of model \eqref{eq:IIR} is denoted as $\mathcal{G} = (V,E)$, where $V$ is the set of all nodes and $E$ is the set of all directed edges in the graph. Node $v_j \in V$ in the graph represents a particular brain network, defined by the ICA procedure, and its dynamic behavior is described by the corresponding ICA times series $w_j(t)$. Thus the cardinality of $V$ equals the total number of ICA time series. A directed edge from $v_i$ to $v_j$ exists, denoted as $e_{ji}\in E$ or $e_{ji}\in \mathcal{G}$, if and only if $G_{ji}$ in \eqref{eq:IIR}, or equivalently, $\theta_{ji}$ in \eqref{eq:ARvecform} is non-zero, which indicates a causal connection from node $v_i$ to $v_j$. We also denote the set of all incoming edges to $v_j$ as $E_j$ or $\mathcal{G}_j$. Now, a graph $\mathcal{G}$ indicates the effective connectivity between brain networks. An example of such a directed graph is shown in Figure \ref{fig:dirgraph}. \par
%
%\begin{figure}[pos=p]
\section{Materials}
\subsection{Simulation data generation}
Before being applied to the real fMRI data, which typically suffers from low sampling rates and measurement noise, the methods for the inference of the effective connectivity should at least show a reasonable performance for the simulated data generated in an ideal setting, i.e. without the above complexities. Since the data generation model is known in simulation, the goal of the simulation study is to test how well the estimated connectivity of one method approximates the true connectivity. \par
In this work, in order to evaluate the relative performances of the Granger-causality analysis and the Bayesian topology identification, we test them in a controlled environment using the simulated data, which is generated by models as defined in \eqref{eq:IIR}. To be precise, we will use the same method for data generation as performed by \citet{BayesShi}. This model type is generally a good approximation of the BOLD signal caused by the neural activity, while the values of coefficients are generated randomly over a general range, which also covers the hemodynamics and the neural activity of brain networks as special cases. \par
The node time series $w_j(t)$ are generated from a model \eqref{eq:IIR}, where some random directed edges $e_{ji}$ are chosen to form a ground truth connectivity $\mathcal{G}_0$, and for each $e_{ji}$ which exists in $\mathcal{G}_0$, its corresponding $G_{ji}(q)$ is a random transfer function \citep{VdHofdynnet}, such that the resulting dynamic network adheres to the IIR model assumptions. The noise $\eta_j(t)$ in \eqref{eq:IIR} is randomly sampled from a Gaussian distribution for each node in the simulation model. Some other related details can be found in Section~\ref{sec:evaluationSimu}.
\subsection{Mozart effect study}
Sixteen volunteers (11 female, 5 male) without a medical history of neurological or psychiatric disease, aged 20-65 (mean 43.3), participated in the experiment. This number of subjects was chosen to be comparable with \citet{MozartCoppola}. To prevent bias, subjects with a variety of taste in music were selected and no subjects were selected who already consistently (2 hours or more per day) listened to Mozart music. Informed consent of the experiment and scans were obtained from each subject. The scanning of healthy volunteers was approved by the Medical Ethical Committee of the Academic Center for Epilepsy Kempenhaeghe (Heeze, The Netherlands).
\begin{figure}
    \centering
    \includegraphics[width=14cm]{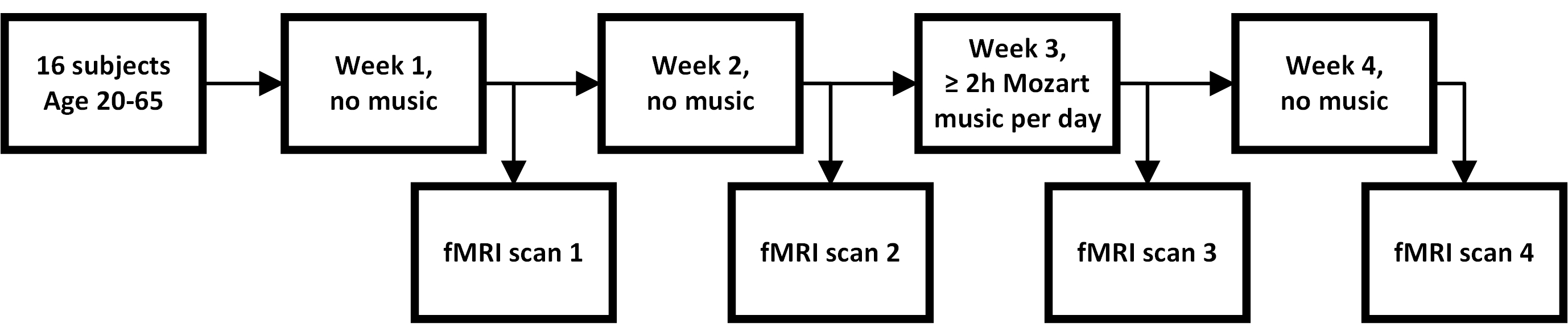}
    \caption{Flowchart of Mozart music experiment. (This is a 1.5 column figure)}
    \label{fig:Mozartflow}
\end{figure}
\begin{figure}
    \centering
    \includegraphics[width=9cm]{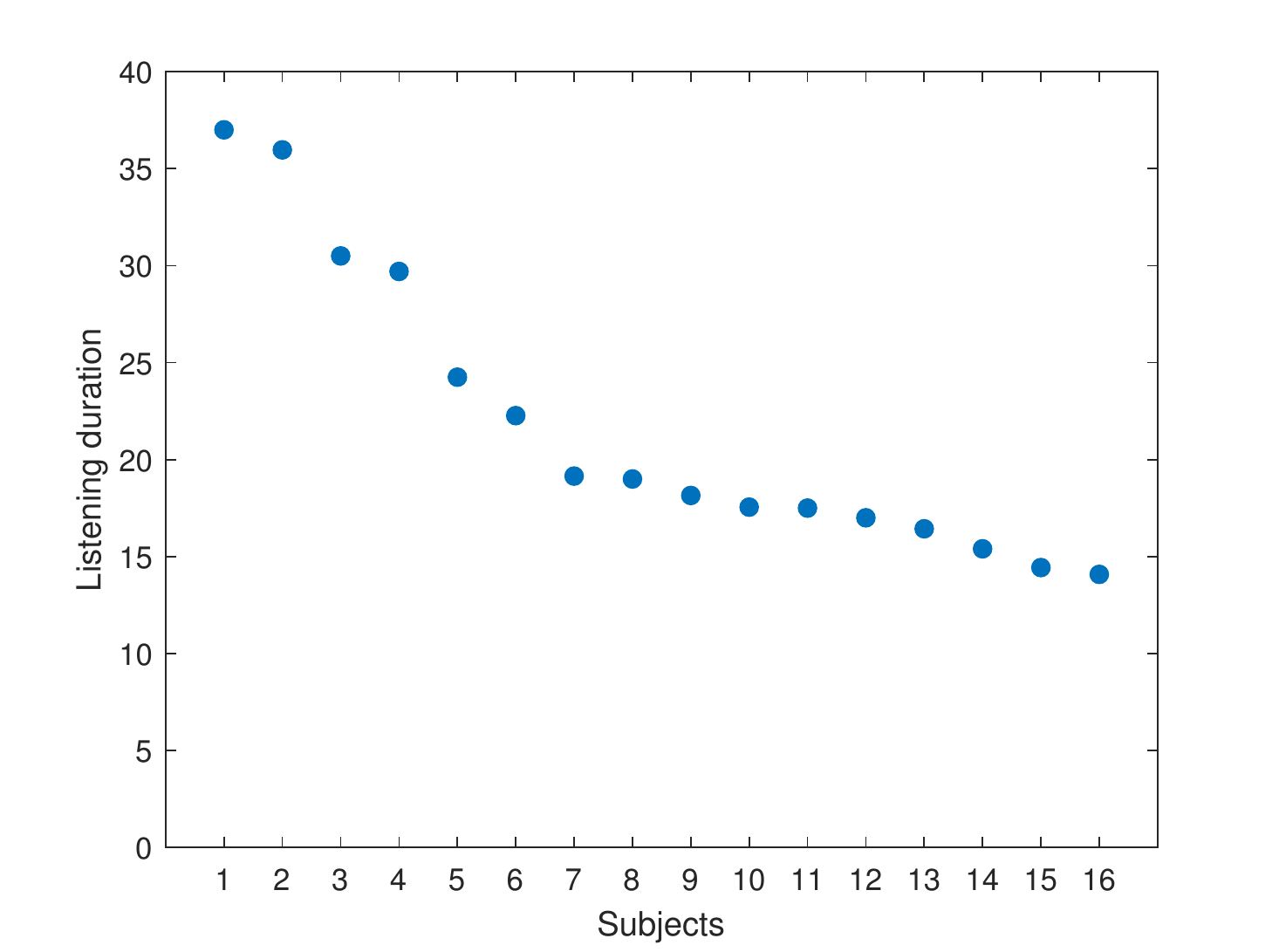}
    \caption{Listening duration in hours of all 16 subjects ordered from the longest to shortest listening subject. (This is a single column figure)}
    \label{fig:lisdur}
\end{figure}
%
%\begin{table}[pos=p]
\begin{table}
    \centering
    \caption{List of all brain networks found in the group ICA decomposition.}
    \begin{tabular*}{\tblwidth}{@{} LLL@{} }
    \toprule
        \# & Shorthand & Description  \\
        \midrule
        1 & DMN\_ANT & default mode network, anterior \\
        2 & MED\_VISU & medial visual cortex \\
        3 & OCC\_LAT & occipetal \& lateral visual cortex \\
        4 & DAN & dorsal attention network \\
        5 & FPR & fronto-parietal right network\\
        6 & SM\_LAT & sensori-motor, lateral network \\
        7 & FPL & fronto-parietal left network\\
        8 & VAN & ventral attention network \\
        9 & LING\_FUS & lingual fusiform cortex \\
        10 & DMN\_POS & default mode network, posterior \\
        11 & N11 & noise \\
        12 & SAL\_AUDI & salience auditory network \\
        13 & ANG & angular gyrus \\
        14 & N14 & White matter \\
        15 & SUPTEMP & superior temporal gyrus \\
        16 & CEN & central executive network \\
        17 & N17 & noise \\
        18 & SM\_SUP & sensori motor medial superior \\
        19 & CEREB & cerebellum \\
        20 & N20 & borders and movements \\
        \bottomrule
    \end{tabular*}
    \label{tab:brainreg}
\end{table}
\subsubsection{Mozart's Sonata K448 experiment}
Four resting-state fMRI scans are collected for each subject in the study, and the experiment procedure is summarized in Figure \ref{fig:Mozartflow}. Each of the four scans are taken one week apart in time. In the week between scan 1 and 2, subjects were instructed not to listen to Mozart's Sonata K448 for any amount of time. Then, the subjects were instructed to listen for a minimum of 14 hours to Mozart's Sonata K448 during the week between scan 2 and 3 and to listen for at least 2 hours each day unless this was not possible due to unavoidable circumstances. This listening duration was chosen to be consistent with \citet{MozartCoppola}, where a similar listening duration was used. The listening duration of each subject during the week of music exposure was recorded by the subjects themselves. In Figure \ref{fig:lisdur} we can see the spread of total listening duration across all 16 subjects. After the third scan subjects were once again instructed to stop listening to Mozart music and then the final scan is taken one week after the subjects stopped listening to Mozart's Sonata K448. From the first scan to the fourth scan, subjects were also instructed to avoid listening to any other music, but limited exposure to music in public places could in some cases not be avoided. Other auditory stimuli, from for example entertainment on television or internet, could also affect brain activity and bias the results. However, the first two scans establish a baseline of regular brain activity and thus should rule out the effect of most regularly present auditory stimuli, when we compare the baseline activity to the brain activity after the subjects listened to Mozart music.
\par
\begin{figure*}
\centering
    \begin{subfigure}[t]{0.09\textwidth}
    \centering
    \caption*{\scriptsize{DMN\_ANT}}
    \vspace{-2mm}
    \includegraphics[width=0.9\textwidth]{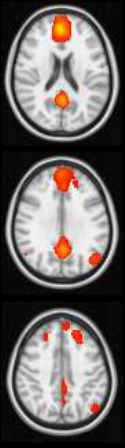}
    \end{subfigure}
    \hfill
    \begin{subfigure}[t]{0.09\textwidth}
    \centering
    \caption*{\scriptsize{MED\_VISU}}
    \vspace{-2mm}
    \includegraphics[width=0.9\textwidth]{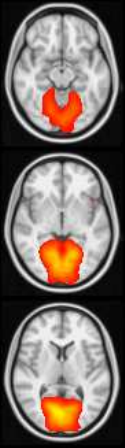} 
    \end{subfigure}
    \hfill
    \begin{subfigure}[t]{0.09\textwidth}
    \caption*{\scriptsize{OCC\_LAT}}
    \vspace{-2mm}
    \centering
    \includegraphics[width=0.9\textwidth]{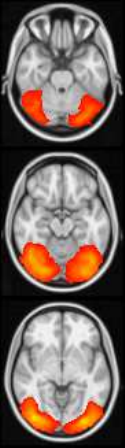} 
    \end{subfigure}
    \hfill
    \begin{subfigure}[t]{0.09\textwidth}
    \centering
    \caption*{\scriptsize{DAN}}
    \vspace{-2mm}
    \includegraphics[width=0.9\textwidth]{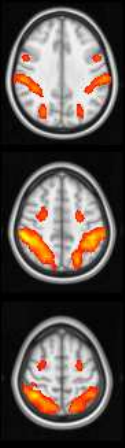} 
    \end{subfigure}
    \hfill
    \begin{subfigure}[t]{0.09\textwidth}
    \centering
    \caption*{\scriptsize{FPR}}
    \vspace{-2mm}
    \includegraphics[width=0.9\textwidth]{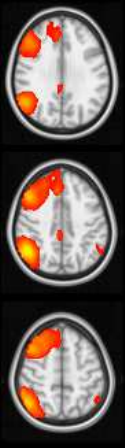} 
    \end{subfigure}
    \hfill
    \begin{subfigure}[t]{0.09\textwidth}
    \centering
    \caption*{\scriptsize{SM\_LAT}}
    \vspace{-2mm}
    \includegraphics[width=0.9\textwidth]{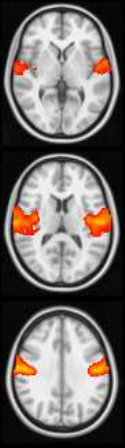} 
    \end{subfigure}
    \hfill
    \begin{subfigure}[t]{0.09\textwidth}
    \centering
    \caption*{\scriptsize{FPL}}
    \vspace{-2mm}
    \includegraphics[width=0.9\textwidth]{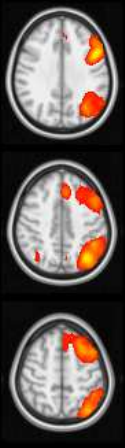} 
    \end{subfigure}
    \hfill
    \begin{subfigure}[t]{0.09\textwidth}
    \centering
    \caption*{\scriptsize{VAN}}
    \vspace{-2mm}
    \includegraphics[width=0.9\textwidth]{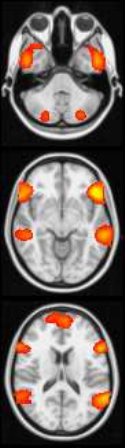} 
    \end{subfigure}
    \hfill
    \begin{subfigure}[t]{0.09\textwidth}
    \centering
    \caption*{\scriptsize{LING\_FUS}}
    \vspace{-2mm}
    \includegraphics[width=0.9\textwidth]{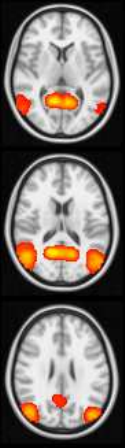} 
    \end{subfigure}
    \hfill
    \begin{subfigure}[t]{0.09\textwidth}
    \centering
    \caption*{\scriptsize{DMN\_POS}}
    \vspace{-2mm}
    \includegraphics[width=0.9\textwidth]{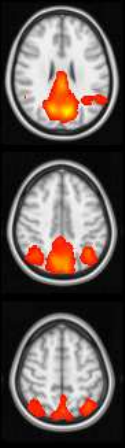} 
    \end{subfigure}
    \hfill
    \begin{subfigure}[t]{0.09\textwidth}
    \centering
    \vspace{2mm}
    \caption*{\scriptsize{N11}}
    \vspace{-2mm}
    \includegraphics[width=0.9\textwidth]{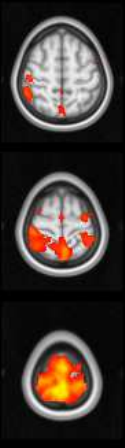} 
    \end{subfigure}
    \hfill
    \begin{subfigure}[t]{0.09\textwidth}
    \centering
    \vspace{2mm}
    \caption*{\scriptsize{SAL\_AUDI}}
    \vspace{-2mm}
    \includegraphics[width=0.9\textwidth]{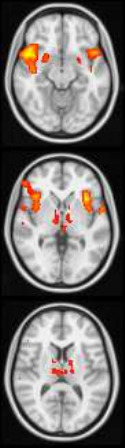} 
    \end{subfigure}
    \hfill
    \begin{subfigure}[t]{0.09\textwidth}
    \centering
    \vspace{2mm}
    \caption*{\scriptsize{ANG}}
    \vspace{-2mm}
    \includegraphics[width=0.9\textwidth]{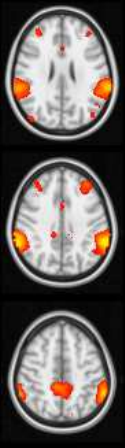} 
    \end{subfigure}
    \hfill
    \begin{subfigure}[t]{0.09\textwidth}
    \centering
    \vspace{2mm}
    \caption*{\scriptsize{N14}}
    \vspace{-2mm}
    \includegraphics[width=0.9\textwidth]{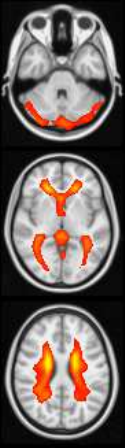} 
    \end{subfigure}
    \hfill
    \begin{subfigure}[t]{0.09\textwidth}
    \centering
    \vspace{2mm}
    \caption*{\scriptsize{SUPTEMP}}
    \vspace{-2mm}
    \includegraphics[width=0.9\textwidth]{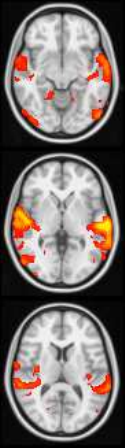} 
    \end{subfigure}
    \hfill
    \begin{subfigure}[t]{0.09\textwidth}
    \centering
    \vspace{2mm}
    \caption*{\scriptsize{CEN}}
    \vspace{-2mm}
    \includegraphics[width=0.9\textwidth]{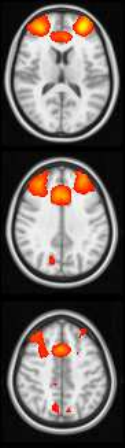} 
    \end{subfigure}
    \hfill
    \begin{subfigure}[t]{0.09\textwidth}
    \centering
    \vspace{2mm}
    \caption*{\scriptsize{N17}}
    \vspace{-2mm}
    \includegraphics[width=0.9\textwidth]{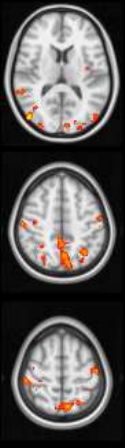} 
    \end{subfigure}
    \hfill
    \begin{subfigure}[t]{0.09\textwidth}
    \centering
    \vspace{2mm}
    \caption*{\scriptsize{SM\_SUP}}
    \vspace{-2mm}
    \includegraphics[width=0.9\textwidth]{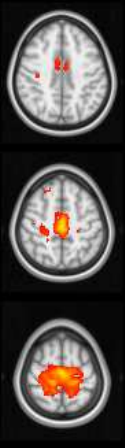} 
    \end{subfigure}
    \hfill
    \begin{subfigure}[t]{0.09\textwidth}
    \centering
    \vspace{2mm}
    \caption*{\scriptsize{CEREB}}
    \vspace{-2mm}
    \includegraphics[width=0.9\textwidth]{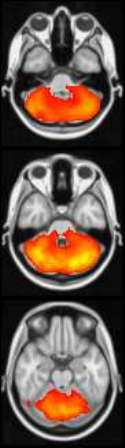} 
    \end{subfigure}
    \hfill
    \begin{subfigure}[t]{0.09\textwidth}
    \centering
    \vspace{2mm}
    \caption*{\scriptsize{N20}}
    \vspace{-2mm}
    \includegraphics[width=0.9\textwidth]{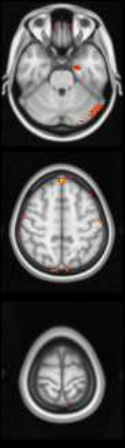} 
    \end{subfigure}
    \caption{Spatial maps of the 20 active brain networks found through the ICA decomposition. Each image consists of 3 relevant horizontal slices of the brain, where the spatial map is indicated by the red color scale. (This is a 2-column figure)}
    \label{fig:spatialmaps}
\end{figure*}
\subsubsection{Acquisition}
Imaging was performed using a 3T Philips Achieva MRI scanner. A T1-weighted reference scan was recorded using a 3D spoiled gradient-echo sequence (T1TFE), TR/TE: 8.3/3.5ms. The matrix size was $240\times240\times180$ with isotropic voxels of 1mm$^3$. Total T1 scan duration was 600s. Second, a multi-echo time fMRI scan was recorded using a gradient-echo EPI sequence with TR/TE/ES: 2000/12/23ms. 26 slices with a width of 4.5mm (0mm gap) were recorded. The data were acquired with $3.5\times3.5$mm in-plane resolution, final in-plane resolution was also $3.5\times3.5$mm ($64\times64$). SENSE \citep{MRISENSE} acceleration was used with SENSE factor of 2.7 in the read-out direction. The total multi-echo scan time was 608s.\par
During all scans the subjects were asked to remain still with their eyes open and to focus on a projected dark-blue cross. Physiology data (heart and respiratory rate) were recorded using a standard scanner set-up.
\subsubsection{Preprocessing}
First, non-BOLD signals caused by physiological effects such as movement, breathing and pulse were removed by multi-echo ICA (MEICA) \citep{MEICAKundu}. Then, using FMRIB Software Library (FSL) \citep{gICA} (v5.0, https://fsl.fmrib.ox.ac.uk/fsl/fslwiki/) all 64 data sets were temporally concatenated and a group ICA was performed to identify 20 independent components, which are listed in Table \ref{tab:brainreg}. The spatial maps of these brain networks can be found in Figure \ref{fig:spatialmaps}. Next, dual regression \citep{gICA} was performed to obtain individual spatial maps and time series for each fMRI scan of the subjects, where each ICA time series consists of $N=300$ data points.
\section{Methods}
This section is divided in two parts. In the first part, we briefly discuss both the Granger-causality analysis and the Bayesian topology identification. Next, we detail how we apply both methods to the simulation data. Then, we develop the extension to the Bayesian method. In the second part, we first validate the use of AR models to model the ICA time series. Then we will apply the extended Bayesian topology identification and the Granger-causality analysis to the ICA time series of the Mozart effect study.
\subsection{Granger-causality analysis}
Here we briefly describe the well-established Granger-causality analysis, which will be used as a reference to compare to the Bayesian topology identification. The Granger-causality analysis is based on the concept that if for a given $j$ and $i$, $\theta_{ji}$ in \eqref{eq:ARvecform} is significantly non-zero, then setting parameter vector $\theta_{ji}$ to 0 will significantly increase the size of the residual $\eta_j(t)$. However to relate this change in the residual to a real connection between two brain networks and not from a shared influence from other regions, we will infer connectivity only using the conditional Granger-causality analysis.\par
The existence of a connection $e_{ji}$ is inferred by fitting two different AR models to the measurement data and comparing the residuals. The first AR model includes all $L$ node time series in the regression as in \eqref{eq:ARvecform}. In the second AR model $A_{ji}$ corresponding to connection $e_{ji}$ is excluded:
\begin{equation}\label{eq:redAR}
    w_j=\sum_{k\in I\backslash i} A_{jk}\cdot \theta_{jk}'+\eta_j'.
\end{equation}
The model order $m$, i.e. the length of parameter vectors $\theta_{ji}$, is selected through the Akaike information criterion (AIC) \citep{AICakaike}, to sufficiently model the interactions in the data without overfitting, as it can make the Granger-causality analysis unreliable \citep{GCtoolbox}.\par
Next, the Granger-value $\mathcal{F}_{ji}$ corresponding to connection $e_{ji}$ is defined as the logarithm of the ratio of the variances of the residuals:
\begin{equation} \label{eq:Gval}
    \mathcal{F}_{ji} = \log\frac{\operatorname{var}[\eta_j']}{\operatorname{var}[\eta_j]}.
\end{equation}
The Granger-value $\mathcal{F}_{ji}$ is the degree to which the addition of connection $e_{ji}$ helps predict $w_j$ and as such, a significantly non-zero $\mathcal{F}_{ji}$ indicates the existence of connection $e_{ji}$. Furthermore, $\mathcal{F}_{ji}$ asymptotically follows a $\chi^2$ distribution \citep{GCtoolbox} as data length $N \rightarrow \infty$ and thus we can perform a statistical test \citep{GCtoolbox} to infer if $\mathcal{F}_{ji}$ is significantly non-zero. Furthermore, the Granger-value is equivalent to the transfer entropy \citep{BarnettTE,KaiserTE,SchreiberTE}, a measurement of the rate of information transfer between two time series, which is an information-theoretic concept. As such, $\mathcal{F}_{ji}$ can be interpreted as a measurement of connectivity strength. Therefore, a change in Granger-value of a connection between two fMRI scans of a subject indicates a difference in the connectivity strength of that connection.\par
However, it is important to note that, when inferring effective connectivity using ICA time series from fMRI measurements, it is ill-advised to infer a graph estimate of the connectivity, as prior research \citep{GCdowns,GCtoolbox} indicates that the statistical tests become unreliable through an overall increase in false positives. Nevertheless, the equivalence of Granger-causality with the transfer-entropy remains valid and therefore we can use a paired t-test to test for differences in connectivity strength between two fMRI scans. \par
For the simulation study, since the effective connectivity of the data-generation systems is known, our goal is to quantify the quality of $\mathcal{F}_{ji}$ as a measurement of connectivity. To be able to compare the quality of the estimates of both methods on simulated data we will use an F-test to determine a graph estimate, as the Bayesian method also infers a graph estimate. If we would use the paired t-test on the simulation data, it would make comparing the estimation performance between the methods impossible. Even though this is not the method we will use on the real data, it still gives us an indication of the reliability of $\mathcal{F}_{ji}$ as a measurement of connectivity. As such, we will infer an estimate $\Hat{\mathcal{G}}$ of the ground truth connectivity $\mathcal{G}_0$ by testing the significance of $\mathcal{F}_{ji}$ for all connections $e_{ji}$ and adding the connection to $\Hat{\mathcal{G}}$ if the statistical test is significant. To calculate the reduced regression in \eqref{eq:redAR}, the Granger-values in \eqref{eq:Gval} and the significance of the Granger-values, we will use the MVGC MATLAB toolbox \citep{GCtoolbox} (version 1.0, https://users.sussex.ac.uk/~lionelb/MVGC/).\par
\subsection{Bayesian topology identification} \label{Bayesianmethod}
Bayesian topology identification \citep{BayesShi,BayesChiuso} is a Bayesian machine learning method that infers from node time series data $D=\{w_1,\ldots,w_L\}$ an estimate $\Hat{\mathcal{G}}$ of $\mathcal{G}_0$, the connectivity of the dynamic network. A Bayesian model selection \citep{MLbishop} approach is used to compare the posterior probability $p(\mathcal{G}|D)$ of different graphs and then selects the graph with maximum $p(\mathcal{G}|D)$ as the optimum graph estimate $\Hat{\mathcal{G}}$. Given a graph, the posterior probability $p(\mathcal{G}|D)$ can be calculated using Bayes' rule:
\begin{equation}
    p(\mathcal{G}|D)=\frac{p(D|\mathcal{G})p(\mathcal{G})}{p(D)}.
\end{equation}
The evidence $p(D)$ can be difficult to calculate, as it requires a marginalization over the set of all graphs. Instead, under the assumption that the prior distribution of graphs is uninformative, i.e. $p(\mathcal{G})$ is constant, the graph with the maximum marginal likelihood $p(D|\mathcal{G})$ will also have the maximum posterior odds $p(\mathcal{G}|D)$. As such, the graph estimate $\Hat{\mathcal{G}}$ is selected as the graph with the maximum marginal likelihood.\par
The marginal likelihood $p(D|\mathcal{G})$ can be computed through marginalization over its parameters $\theta$:
\begin{equation}\label{eq:margll}
    p(D|\mathcal{G})=\int p(D|\theta,\mathcal{G}) p(\theta|\mathcal{G}) d\theta,
\end{equation}
where $p(D|\theta,\mathcal{G})$ is the likelihood of model parameters $\theta$ and a graph $\mathcal{G}$ and $p(\theta|\mathcal{G})$ is the prior distribution of $\theta$ given a graph $\mathcal{G}$. \par
In the context of this study, the model parameters $\theta$ in the parameter prior $p(\theta|\mathcal{G})$ indicate the set of all $\theta_{ji}$ of an AR model as defined in \eqref{eq:ARvecform}, where parameter vectors $\theta_{ji}$ are modelled as Gaussian random vectors. The co-variance matrices of these random vectors are parametrized with some hyperparameters, which encode the assumption of stability. The hyperparameters of random parameter vectors $\theta_{ji}$ and the variance $\sigma_j^2$ of $\eta_j$ are estimated using an Expectation-Maximization (EM) algorithm \citep{MLbishop} as described in \citep{BayesShi}. Because the $\theta_{ji}$ are modeled as random vectors, we can simply set model order $m$ in \eqref{eq:ARvecform} to some large number at most equal to the data length $N$, instead of determining an optimal model order, as the estimated hyperparameters will determine the relevancy of each of the $m$ parameters.\par
Now, under the assumption of Gaussian distributed noise, $p(D|\theta,\mathcal{G})$ can be computed based on \eqref{eq:ARvecform}:
\begin{equation}
    p(D|\theta,\mathcal{G})=\prod_{j=1}^{L}
    \mathcal{N}\left(\sum_{i\in \operatorname{C}} 
    A_{ji}\theta_{ji},\ 
    \sigma_j^2I_N\right),
\end{equation}
where $C={k|e_{jk}\in \mathcal{G}}$. Given this parametrization, the integral in \eqref{eq:margll} has a closed form solution \citep{BayesShi} and thus we can avoid a numerical integration, which could be computationally costly.\par
The graph $\mathcal{G}$ with the maximum marginal likelihood $p(D|\mathcal{G})$ is chosen as the connectivity estimate $\Hat{\mathcal{G}}$. This maximum can be found by comparison of the marginal likelihoods of all possible graphs. To avoid the combinatorial problem of comparison of all possible graphs, a greedy search algorithm \citep{BayesShi,BayesSearch} is employed that efficiently finds a graph estimate $\Hat{\mathcal{G}}$.\par
To perform the Bayesian topology identification, the implementation from \citet{BayesShi} is used (version 1.0, https://codeocean.com/capsule/3224411/tree/v1).
\subsection{Evaluation of methods in simulation}\label{sec:evaluationSimu}
To motivate the use of the Bayesian topology identification for the inference of brain network connectivity, we will evaluate whether the Bayesian approach has any advantages over Granger-causality analysis in the estimation of the connectivity using simulation data. Now, recall that the true effective connectivity of the simulation model $\mathcal{G}_0$ is known. The graph estimates $\Hat{\mathcal{G}}$ of both methods are compared to $\mathcal{G}_0$ and the quality of the estimate is quantified using the true positive rate (TPR) and the false positive rate (FPR):
\begin{equation}\label{eq:perfcrit}
\begin{aligned}
    &\operatorname{TPR}=\operatorname{TP} / \operatorname{P},\\
    &\operatorname{FPR}=\operatorname{FP} / \operatorname{F},
\end{aligned}
\end{equation}
where the TPR is the number of edges in $\Hat{\mathcal{G}}$ that also exist in $\mathcal{G}_0$, denoted as TP, over the total number of edges in $\mathcal{G}_0$, denoted as P. The FPR is the number of edges in $\Hat{\mathcal{G}}$ that do not exist in $\mathcal{G}_0$, denoted as FP, divided by the total number of edges that do not exist in $\mathcal{G}_0$, denoted as F. To avoid uncertainties in the conclusion on the relative performance of the two methods given certain design parameters for the simulation, i.e. a particular data length $N$ and certain network size in terms of number of nodes, 50 random simulation models with random $\mathcal{G}_0$ are generated, and then TPR and FPR are calculated by averaging the results of the 50 estimates. The random simulation models are generated by selecting random parameters for each transfer for which the connection exists in $\mathcal{G}_0$, such that the transfer is stable. This process is repeated until $(I-G(q))^{-1}$ is stable as per our model assumptions on the model in equation \eqref{eq:IIRfull}.
\par
The design parameters for the simulation are chosen as follows. First, to test how the performance of both methods varies with the data length, we consider three different settings with $N=\{50,\ 300,\ 2000\}$ for six-node dynamic networks. Second, to evaluate how the number of nodes in the network affects the estimation performance, we compare the TPR and FPR of the methods between six and twelve node networks, with $N = 2000$. Third, for each of the previous evaluations we will assess the performance of the Granger-causality analysis over a range of thresholds, determined by a range of significance levels $\alpha=\{0.005,\ 0.05,\ldots,\ 0.95\}$ of the F-test, which are corrected for multiple comparisons using the false discovery rate (FDR) correction \citep{multcomp} in the MVGC MATLAB toolbox. This latter evaluation is performed to rule out differences in the performance of Granger-causality analysis and the Bayesian topology identification caused by a specific choice of significance threshold. Here we also note that the Bayesian method is a point estimate, so no such variation in threshold can or needs to be performed for the Bayesian method. Finally, as we mentioned in Section \ref{Bayesianmethod}, the model order $m$ of the Bayesian method should be chosen to be some large number at most $N$. In practice we set it to $m = \{50, 100, 100\}$ for $N=\{50,\ 300,\ 2000\}$ respectively, as higher $m$ does not visibly increase performance.
\subsection{Bayesian group hypothesis test}
\begin{figure}
    \centering
    \includegraphics[width=14cm]{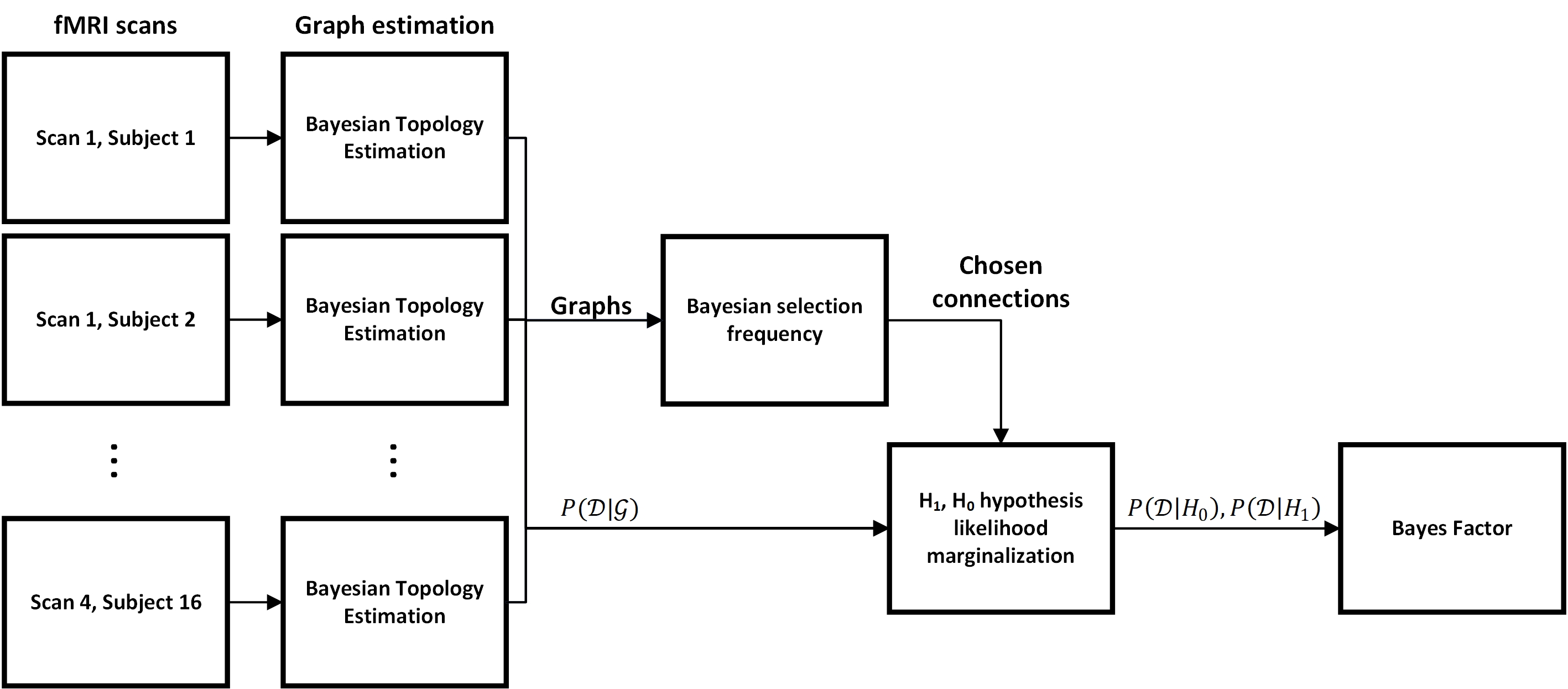}
    \caption{In this figure the procedure of the extended Bayesian method is summarized. (This is a 1.5-column figure)}
    \label{fig:Bayesflow}
\end{figure}
The Bayesian topology identification method is designed to obtain a graph estimate of the effective connectivity from a single fMRI scan, which can be used in simulation to compare with the true connectivity. However, it is not directly applicable to the Mozart effect study using data sets from groups of subjects. In the Mozart effect study there are $4$ groups of data sets of subjects, where each group refers to one of the four measurement moments in the four consecutive weeks. In this latter study, we want to infer how the overall brain network connectivity of the subjects changes over the weeks. Using the Bayesian method, changes in effective connectivity can only be measured through a change in the presence of a connection in the graphs of the subjects. For a group of subjects we define for each connection separately the following hypotheses:  
\begin{equation}
    \begin{aligned}
    &{H}_{1}:  & &\text{Connection}\ e_{ji}\ \text{is present in the group}.\\
    &{H}_{0}: & &\text{Connection}\ e_{ji}\ \text{is absent in the group}.
    \end{aligned}
\end{equation}
To infer if changes in the overall brain network connectivity occurred over the four weeks, we first find the most likely hypothesis for each week separately and then the optimal hypothesis of change is the collection of the most likely hypotheses of each week. If the group hypothesis changes between the weekly scans, this indicates an overall change in the effective connectivity of this connection.\par
The procedure is summarized in Figure~\ref{fig:Bayesflow}, and the preliminary step to asses which hypothesis might be more likely than the other for each week, is the Bayesian selection frequency of the connection, i.e. the number of subjects whose graph estimate includes this connection. It will allow us to choose connections in the Mozart effect study which might be of interested, which is useful, because the analysis of all $380$ possible connections in the Mozart study is estimated to take in the order of weeks or months to complete depending on the used hardware, while many of those connections will not be of interest. If the selection frequency for a given week is close to 0, we expect the most likely hypothesis to be $H_0$, if the selection frequency is closer to 16, i.e. the connection was present in almost all the subjects, we expect the most likely hypothesis will be $H_1$. The Bayesian selection frequency can be easily reported for every connection from the graph estimates of each subject over the four weeks. \par
Now, to actually evaluate which of the two hypotheses is actually more likely than the other for a given connection in one week, we will calculate the likelihood of the two hypotheses:
\begin{equation}\label{eq:hypLL}
\begin{aligned}
    &p(\mathcal{D}|H_{1})= \prod_{k\in S} p({D}^k|e_{ji}),\\
    &p(\mathcal{D}|{H}_{0})= \prod_{k\in S} p({D}^k|\neg e_{ji}),
\end{aligned}
\end{equation}
where with some abuse of notation $e_{ji}$ indicates the existence of the connection, $D^k$ indicates the data from subject $k$, $S$ is the index set of all subjects and $\mathcal{D}$ is defined as the collection of data $D^k$ from each subject in the group. The likelihoods $p({D}^k|e_{ji})$, $p({D}^k|\neg e_{ji})$ in \eqref{eq:hypLL} of each subject are not yet known and must be calculated from the marginal likelihood in \eqref{eq:margll}, which is calculated by the Bayesian method. The likelihoods in \eqref{eq:hypLL} can be calculated through the marginalization of $p(D|\mathcal{G})$ over all graphs for which the hypothesis is true. As an example we use hypothesis $H_1$ here, but the calculation is similar for $H_0$: 
\begin{equation}\label{eq:graphmargll}
    p(D^k|e_{ji})= \sum_{\mathcal{G} \in \mathcal{P}_{1}} p(\mathcal{G}) p(D^k|\mathcal{G}),
\end{equation}
where
\begin{equation}
\begin{aligned}\label{eq:graphset}
    &\mathcal{P}_{1} &=\{\mathcal{G}|e_{ji} \in \mathcal{G}\},\\
    &\mathcal{P}_{0}&=\{\mathcal{G}|e_{ji} \notin \mathcal{G}\}.
\end{aligned}
\end{equation}
This marginalization over graphs can be seen as averaging out the effects that other connections have on the likelihood of the connection in the hypothesis. The calculation of $p(D^k|e_{ji})$ in \eqref{eq:graphmargll} can be further simplified. The simplification of the marginalization is detailed in Appendix \ref{appendA}.\par
Now, we can calculate the hypothesis likelihoods in \eqref{eq:hypLL} and we can find the optimal hypothesis by comparison of the likelihoods of the two hypotheses:
\begin{equation}\label{eq:BF}
    \operatorname{BF}=  2\log\frac{p(\mathcal{D}|H_{1})}{p(\mathcal{D}|H_{0})}.
\end{equation}
This log-likelihood ratio is called the Bayes factor \citep{kassraftBF}, which represents the strength of evidence of one hypothesis against the other. In \citep{kassraftBF} a scale is proposed, which we will use to interpret the size of the Bayes factor. Now, if BF is larger than $0$ then the optimal hypothesis is $H_{1}$ and connection $e_{ji}$ is present in the group. If BF is smaller than $0$ than the alternative hypothesis ${H}_{0}$ is more likely and therefore the connection is absent in the group.\par
Finally, from the likelihood of an hypothesis, we can also calculate the posterior probability of this hypothesis, for example for $H_{1}$:
\begin{equation}
    p(H_{1}|\mathcal{D})=\frac{p(\mathcal{D}|H_{1})p(H_{1})}{p(\mathcal{D})},
\end{equation}
where, because there are only two hypotheses and we choose $p(H_{1})=p(H_{0})=0.5$,
\begin{equation}
    p(\mathcal{D})=p(\mathcal{D}|H_{1})p(H_{1})+p(\mathcal{D}|H_{0})p(H_{0}).
\end{equation}
The posterior probability does not influence the choice of optimal hypothesis chosen using the Bayes factor, but does provide a clearer and more easily interpretable measurement of the strength of evidence in favor of the hypothesis. The larger the posterior probability is, the stronger the evidence that the optimal hypothesis is true given the measurements.\par
\subsection{Validation of ICA time series AR models}
Both Granger-causality analysis and the Bayesian topology identification rely on the assumption that the data is approximately generated by an AR model as defined in \eqref{eq:ARvecform}. If the dynamics of the ICA time series cannot be modeled sufficiently, the two methods becomes less reliable for the inference of brain network connectivity. To assess the goodness of the model fit, we estimate an AR model using all 20 ICA time series from one fMRI scan for each of the 64 scans in the study. The model parameters are calculated using ordinary least squares. Then, we perform a whiteness (auto-correlation) test \citep{SysidLjung} of the residuals of each VAR model, to test whether the AR models can sufficiently model the dynamics in the ICA time series. For model order $m=3$, 96.5\% of the AR model residuals are white, which increases to 99.4\% for $m=5$. Therefore, the results of the tests are satisfactory.
\subsection{Inference of the existence of a Mozart effect}
In our search for a Mozart effect, we will first choose connections with potential effects using the Bayesian selection frequency. Then we will apply the extended Bayesian method and the Granger-causality analysis to the ICA time series of the chosen connections. Based on other studies of the Mozart effect \citep{MozartEEG,MozartRauscher,MozartBodner,MozartYang}, we hypothesize that there will be changes in the effective connectivity from and to brain networks involved in cognitive processing. Furthermore, because the subjects are listening to music we expect to find changes in connectivity between brain networks involved in auditory processing and possibly motor regions.\par
In the Mozart effect study, the scans in week 1 and 2 are used to infer the natural variability in brain network connectivity of the subjects. If a connection has a low variability in effective connectivity in these weeks, it is more likely that any changes in connectivity of subjects between week 2 and week 3 is due to listening to Mozart music. Finally, for the connections of which the connectivity changed between week 2 and 3, we can compare their connectivity of week 3 and 4 to see if the effect lasts even into week 4 or if it was only of short duration.\par
Now, given the criteria described above, we define for each method when a change in effective connectivity took place as a result of listening to Mozart music. For the Bayesian hypothesis test we define the optimal hypotheses for each week such that the collected hypotheses over all four weeks indicate a change in effective connectivity caused by Mozart music. For week 1 and 2 the optimal hypotheses should be the same for both weeks. Then in week 3, the optimal hypothesis should be the opposite of week 1 and 2. Then, the optimal hypothesis in week 4 indicates whether the change between week 2 and 3 lasted into week 4 or not. Finally, in the Granger-causality analysis a significant difference in Granger-values of subjects should only be found between week 2 and 3, and possibly between week 3 and 4, depending on if the change lasted into week 4 or not. 
\subsection{Inference based on listening duration}
We will perform one more analysis of the data, by dividing the subjects' weekly scans into two subgroups based on their listening duration of Mozart music between week 2 and 3. We rank the subjects based on listening time and choose the 8 longest listeners as the first subgroup and the remaining 8 subjects form the second subgroup. The first subgroup of longer listeners listened on average 27:14$\pm$7:07 hours to Mozart music and the second subgroup 16:19$\pm$1:31 hours on average. We perform both the Bayesian hypothesis test and the paired t-test of the Granger-values for both subgroups to assess if effects are influenced by the listening duration of the subjects.\par
\subsection{Data and code availability}
The data supporting the findings of this study can be made available upon request, with a formal data-sharing agreement. The developed code and the simulation data of this study have been made available online \citep{simdatacode}.
\section{Results}
In this section, we will first evaluate the performance of the Bayesian topology identification against Granger-causality analysis on simulation data. Then we apply both methods to the Mozart music ICA time series to infer changes in connectivity caused by listening to Mozart music. Finally, we apply our methods on two subgroups of 8 subjects, divided based on listening duration, to assess if listening duration affects the possible detection of effects.\par
\subsection{Evaluation of methods in simulation}
\begin{figure*}
\centering
    \begin{subfigure}[t]{0.475\textwidth}
    \centering
    \includegraphics[width=0.9\textwidth]{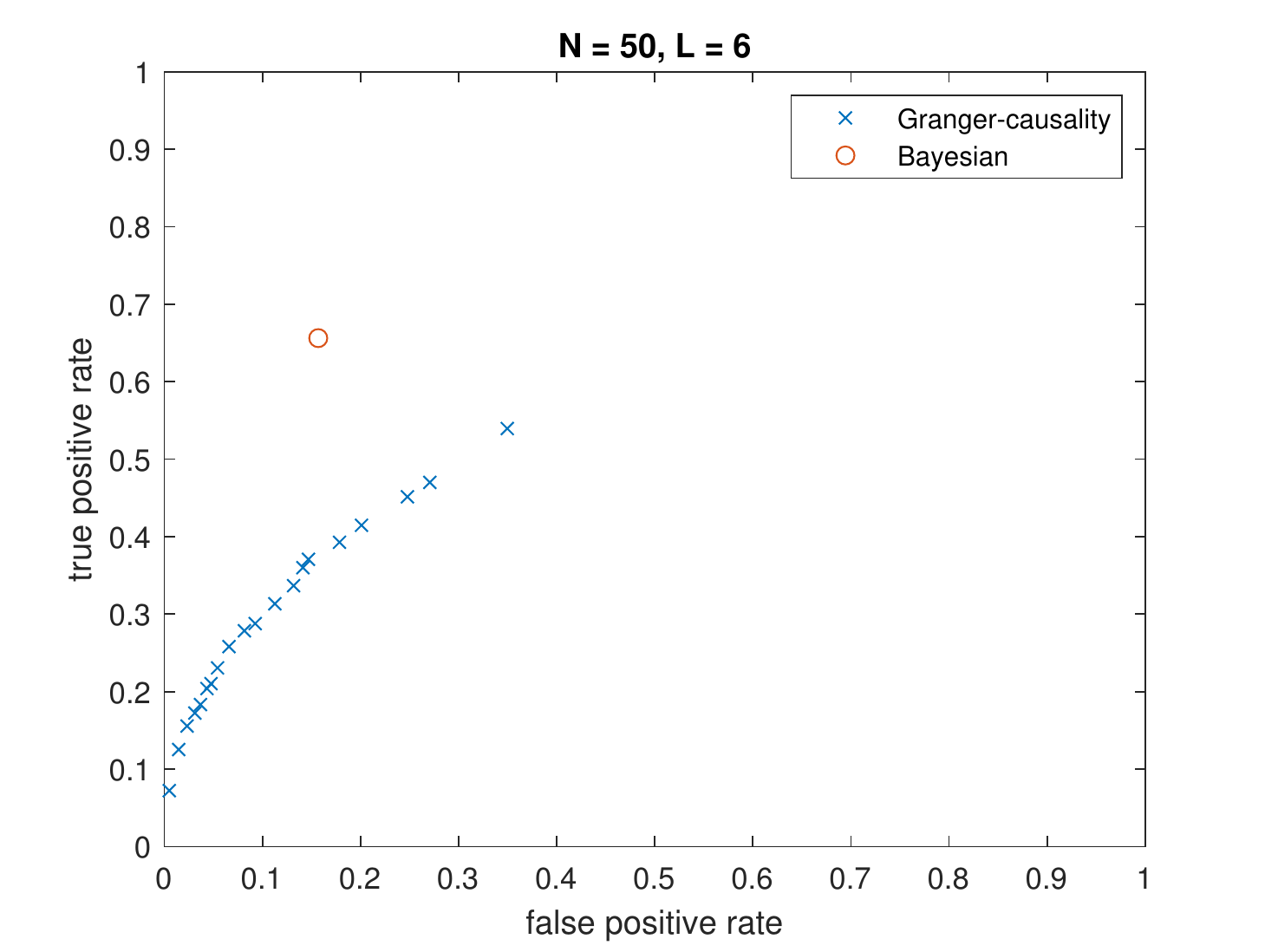} 
    \caption{TPR vs FPR for $\alpha=\{0.005,\ 0.05,\ldots,\ 0.95\}$, $N=50$ and $L=6$.}
    \label{fig:sumimperf1}
    \end{subfigure}
    \hfill
    \begin{subfigure}[t]{0.475\textwidth}
    \centering
    \includegraphics[width=0.9\textwidth]{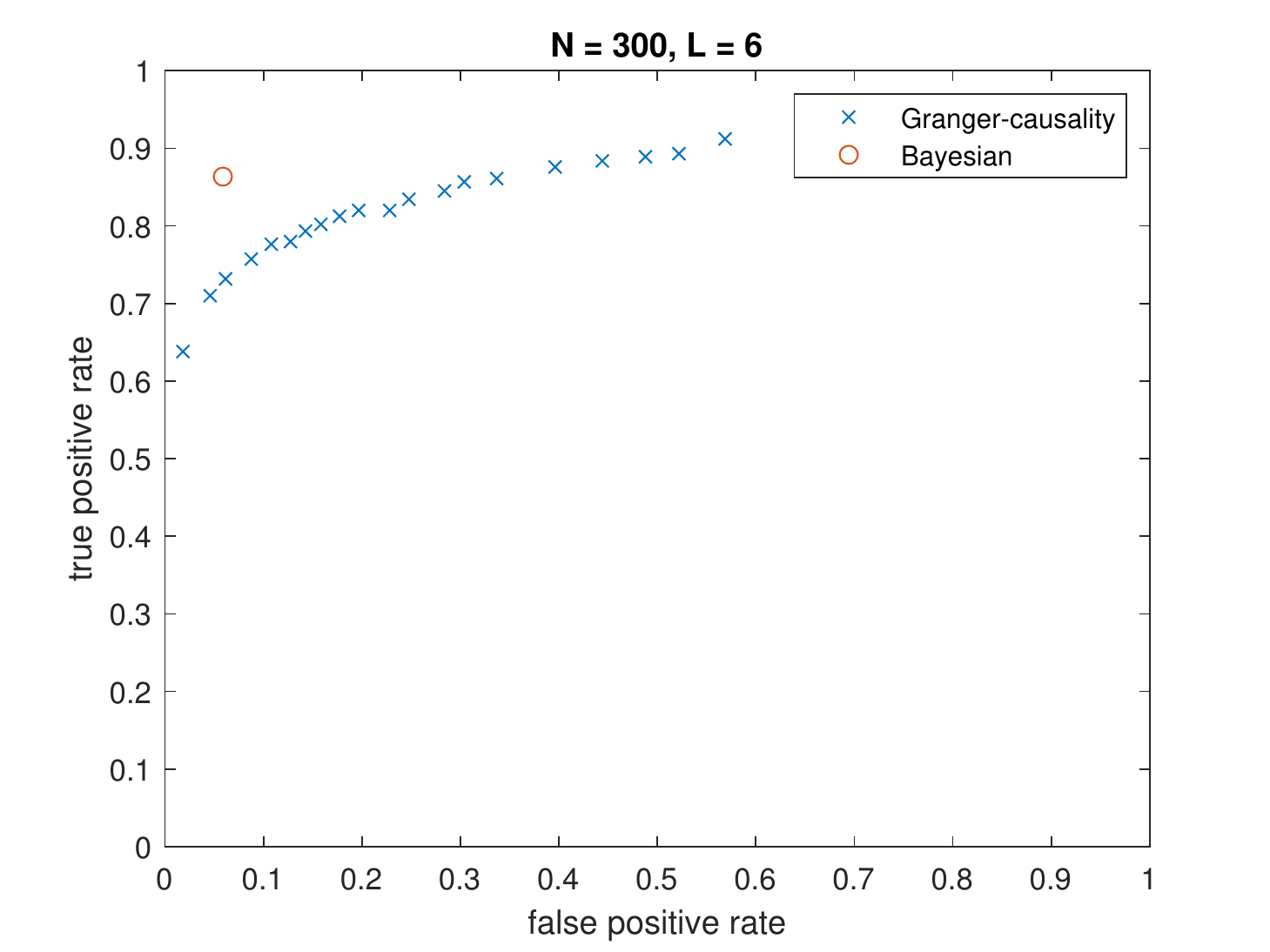}
    \caption{TPR vs FPR for $\alpha=\{0.005,\ 0.05,\ldots,\ 0.95\}$, $N=300$ and $L=6$.}
    \label{fig:sumimperf2}
    \end{subfigure}
    
    \begin{subfigure}[t]{0.475\textwidth}
    \centering
    \includegraphics[width=0.9\linewidth]{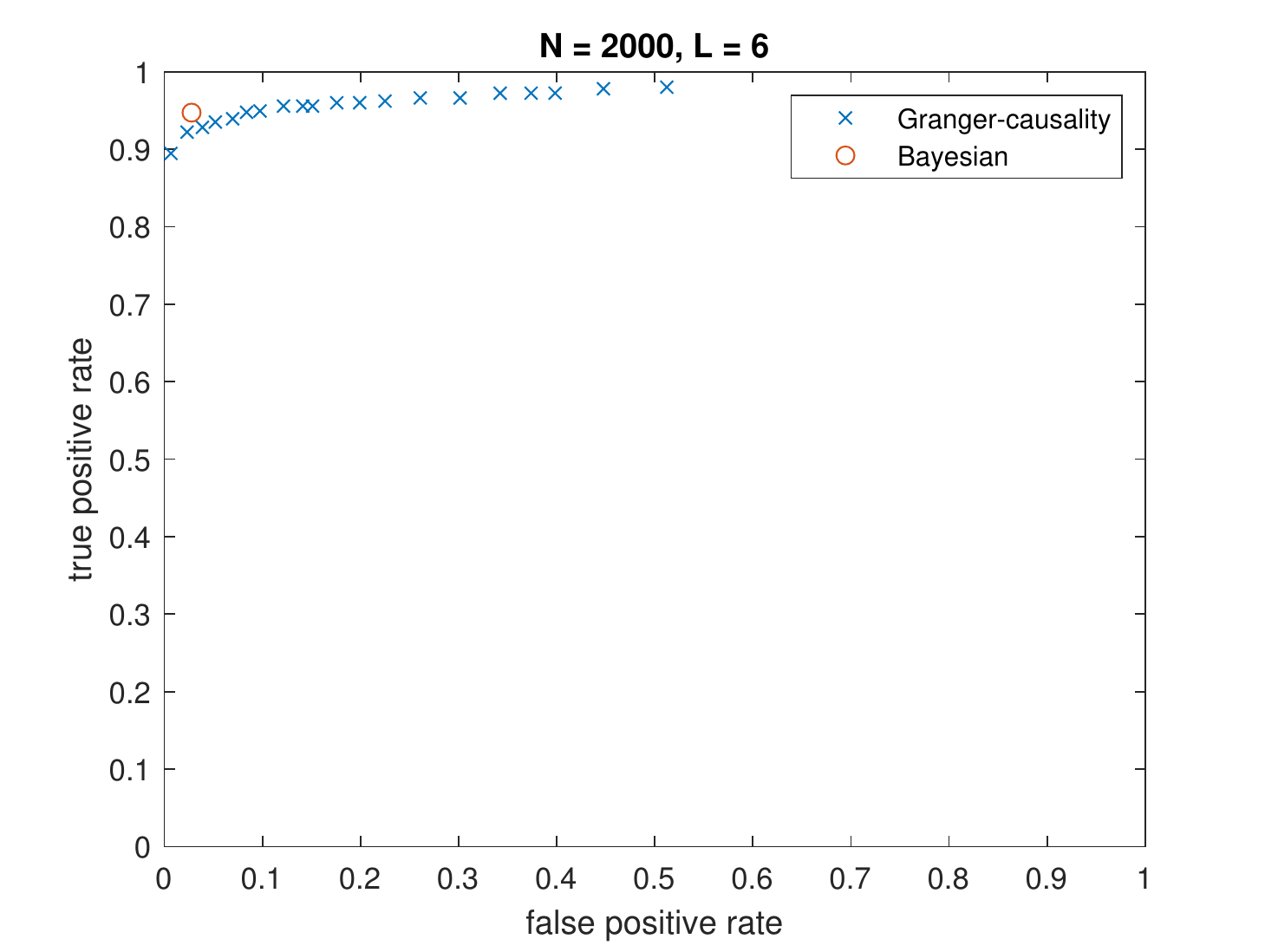}
    \caption{TPR vs FPR for $\alpha=\{0.005,\ 0.05,\ldots,\ 0.95\}$, $N=2000$ and $L=6$.}
    \label{fig:sumimperf3}
    \end{subfigure}
    \hfill
    \begin{subfigure}[t]{0.475\textwidth}
    \centering
    \includegraphics[width=0.9\linewidth]{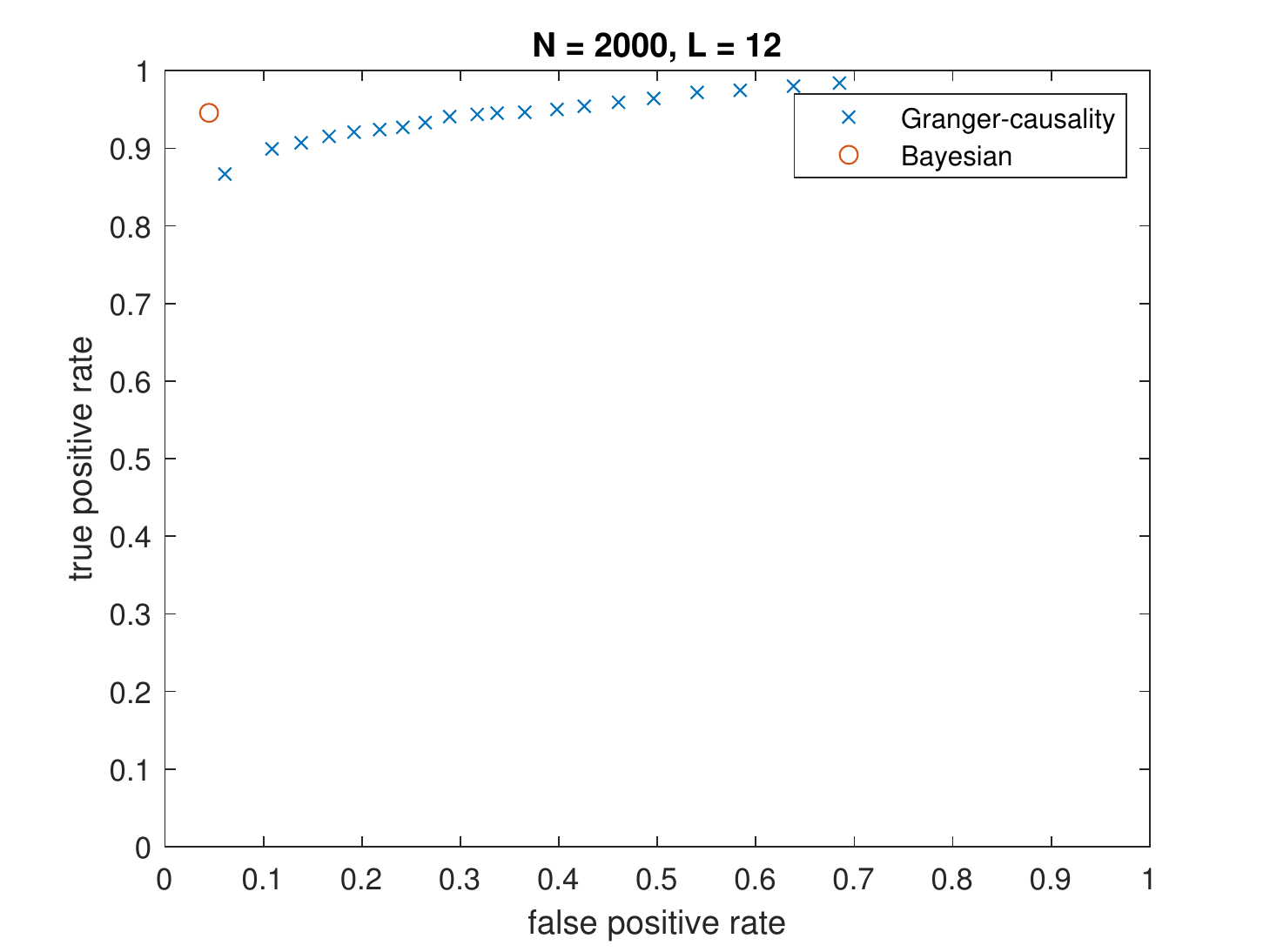}
    \caption{TPR vs FPR for $\alpha=\{0.005,\ 0.05,\ldots,\ 0.95\}$ $N=2000$ and $L=12$.}
    \label{fig:sumimperf4}
    \end{subfigure}
    
    \caption{Average FPR vs TPR of the Bayesian method and Granger-causality analysis graph estimates over $50$ different graphs. One blue cross indicates the TPR and FPR of the Granger-causality analysis for a given threshold. As $\alpha$ increases the TPR and FPR of the Granger-causality analysis increases. The TPR and FPR of the Bayesian topology identification are indicated with a red circle. This is a point estimate as the Bayesian method does not rely on a threshold to determine the existence of connections. (This is a 2-column figure)}
    \label{fig:simperf}
\end{figure*}
The optimal performance in Figures \ref{fig:simperf}a-d is where TPR $=1$ and FPR $=0$, because it implies $\Hat{\mathcal{G}}=\mathcal{G}_0$. The closer we are to this point the better the performance of the method. In Figures \ref{fig:simperf}a-c we see that the Bayesian method shows improvement as the data length increases. It is important to note that the Bayesian method is a point estimate and does not rely on thresholding like the Granger-method. Next, we observe that the Granger method performance improves as the data length increases, where we note that the optimal significance threshold differs between the figures.
Finally, it is clearly visible that in all of Figures \ref{fig:sumimperf1}-\ref{fig:sumimperf3} the Bayesian performance is always closer to the optimal point than the performance of the Granger method for any threshold.\par
From Figures \ref{fig:simperf}a-c we notice that the Bayesian method outperforms the Granger-causality analysis for all three data lengths over all different thresholds. When com\-paring Figures \ref{fig:sumimperf3} and \ref{fig:sumimperf4} we see that there is barely any change in performance for the Bayesian method when the number of nodes in the dynamic network is increased as the TPR is the same in both figures and the FPR is only slightly larger for the twelve-node networks. However, for the Granger-causality analysis, the TPR of almost all thresholds decreases. Furthermore, we also notice a clear increase in the FPR of the Granger-causality analysis for all thresholds. Lastly, one more important detail is that when we inspect Figures \ref{fig:simperf}a-c for the Granger-causality analysis, the optimal performance for each of the figures does not belong to the same threshold, instead it depends on the data length $N$. But based on the change in performance for each threshold as $N$ becomes large, it appears the 5\% threshold will be the optimal threshold for large $N$.\par
\begin{figure*}
\centering
    \begin{subfigure}[t]{0.32\textwidth}
    \centering
    \includegraphics[width=\textwidth]{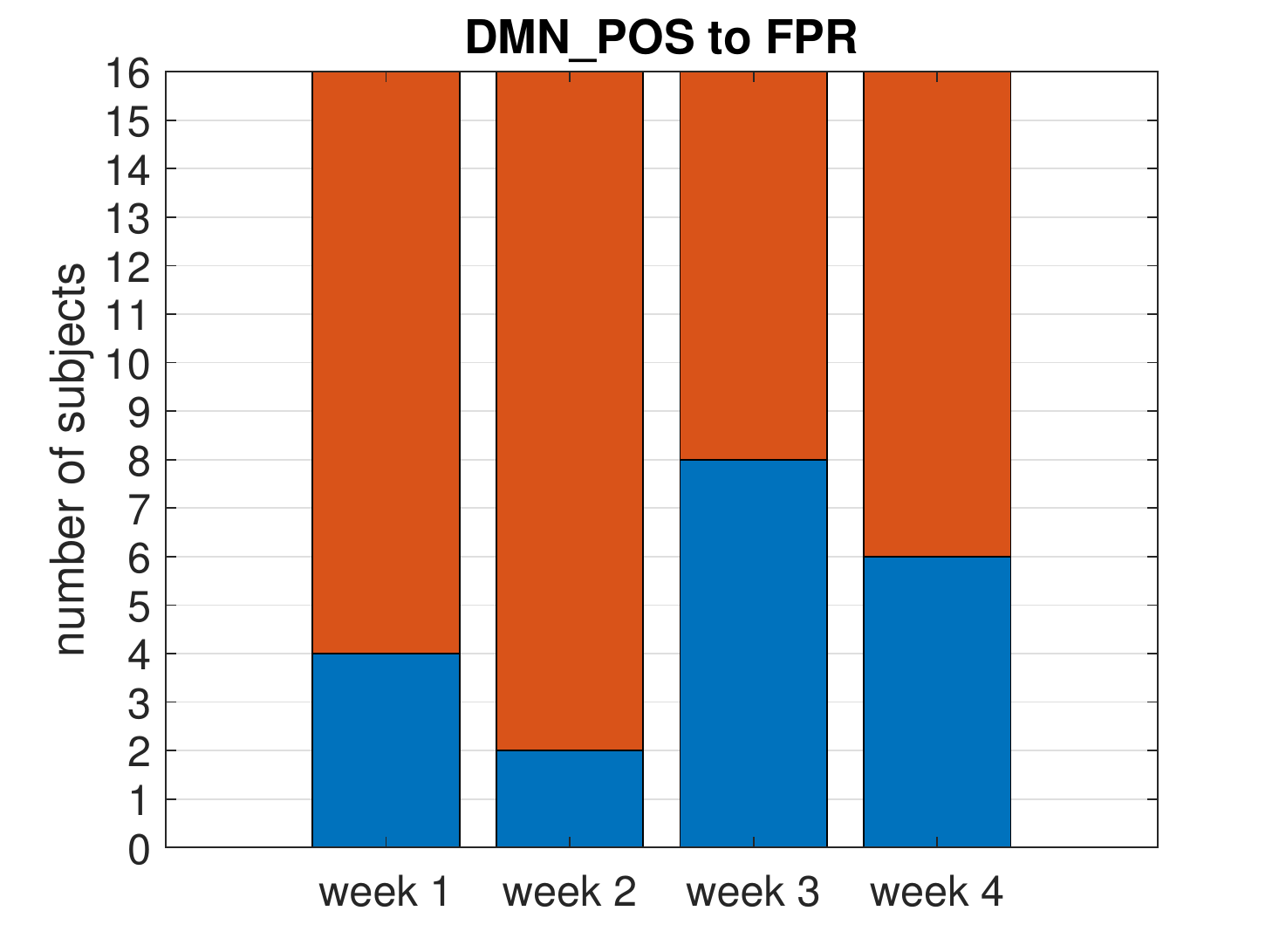}
    \caption{Connection from the posterior default mode network to the fronto-parietal right network.}
    \label{fig:selfr1}
    \end{subfigure}
    \hfill
    \begin{subfigure}[t]{0.32\textwidth}
    \centering
    \includegraphics[width=\textwidth]{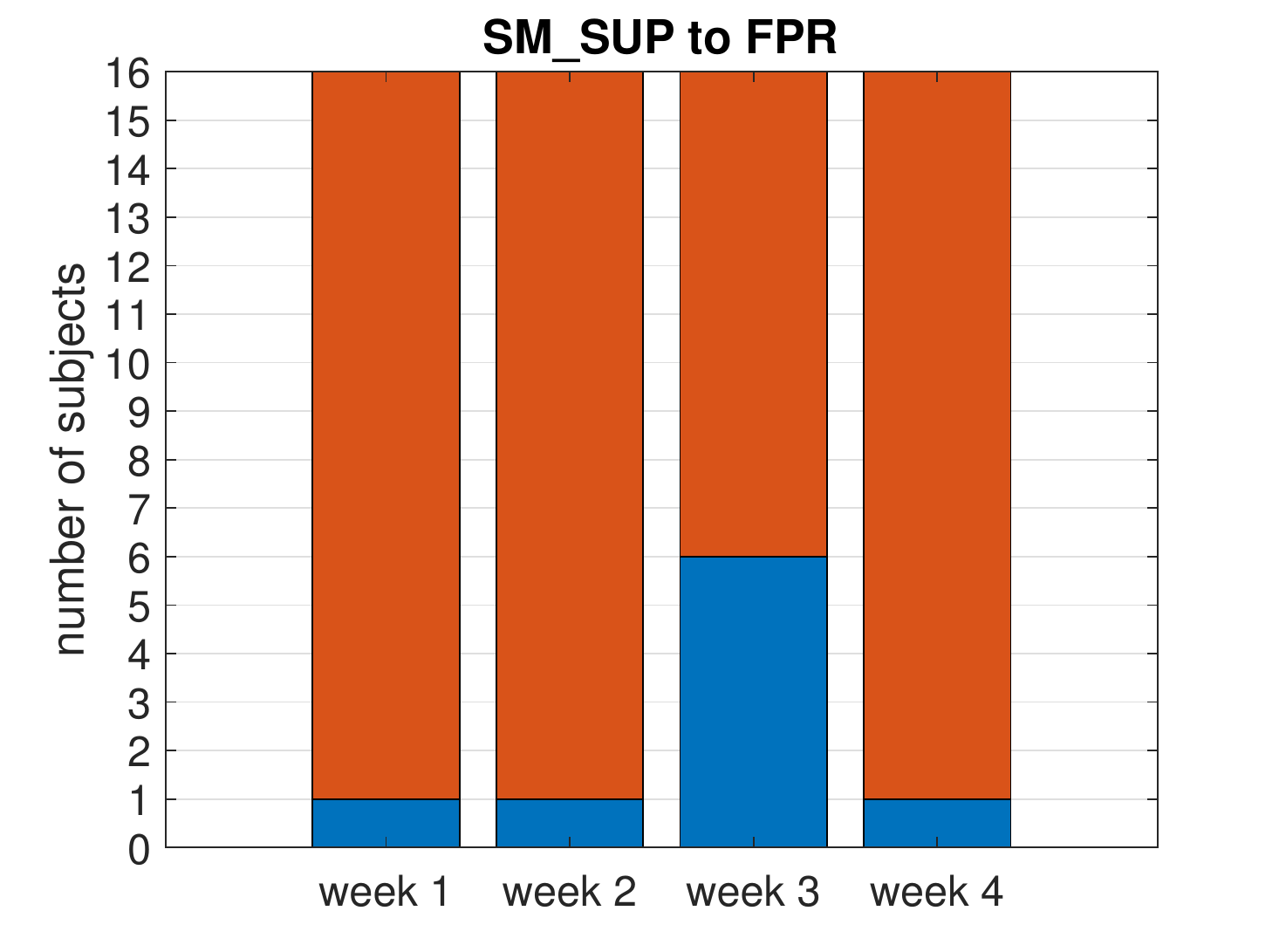}
    \caption{Connection from the sensori-motor superior network to the fronto-parietal right network.}
    \label{fig:selfr2}
    \end{subfigure}
    \hfill
    \begin{subfigure}[t]{0.32\textwidth}
    \centering
    \includegraphics[width=\textwidth]{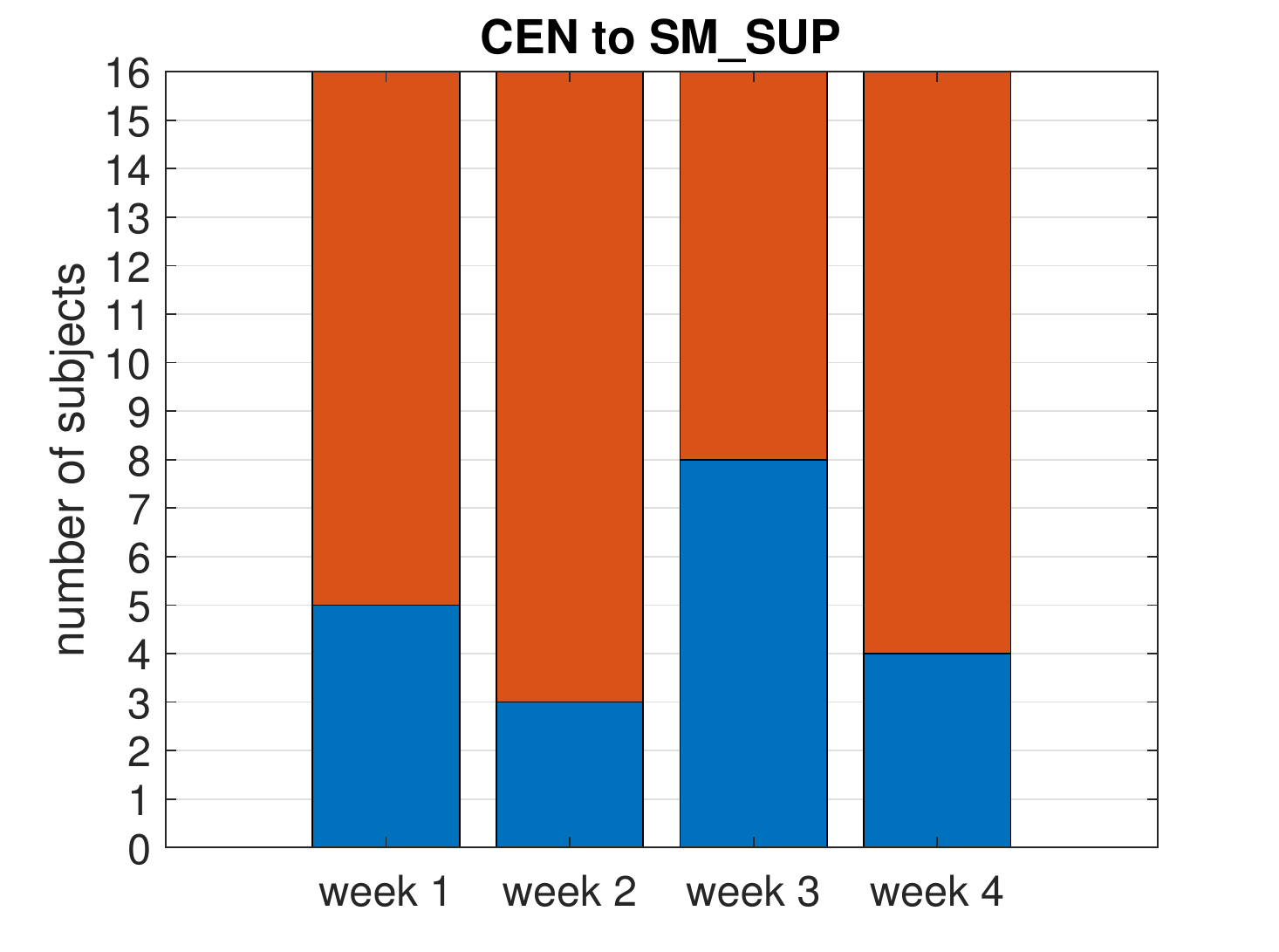}
    \caption{Connection from the central executive network to the sensori-motor superior network.}
    \label{fig:selfr5}
    \end{subfigure}
    
    \medskip
    \begin{subfigure}[t]{0.32\textwidth}
    \centering
    \includegraphics[width=\textwidth]{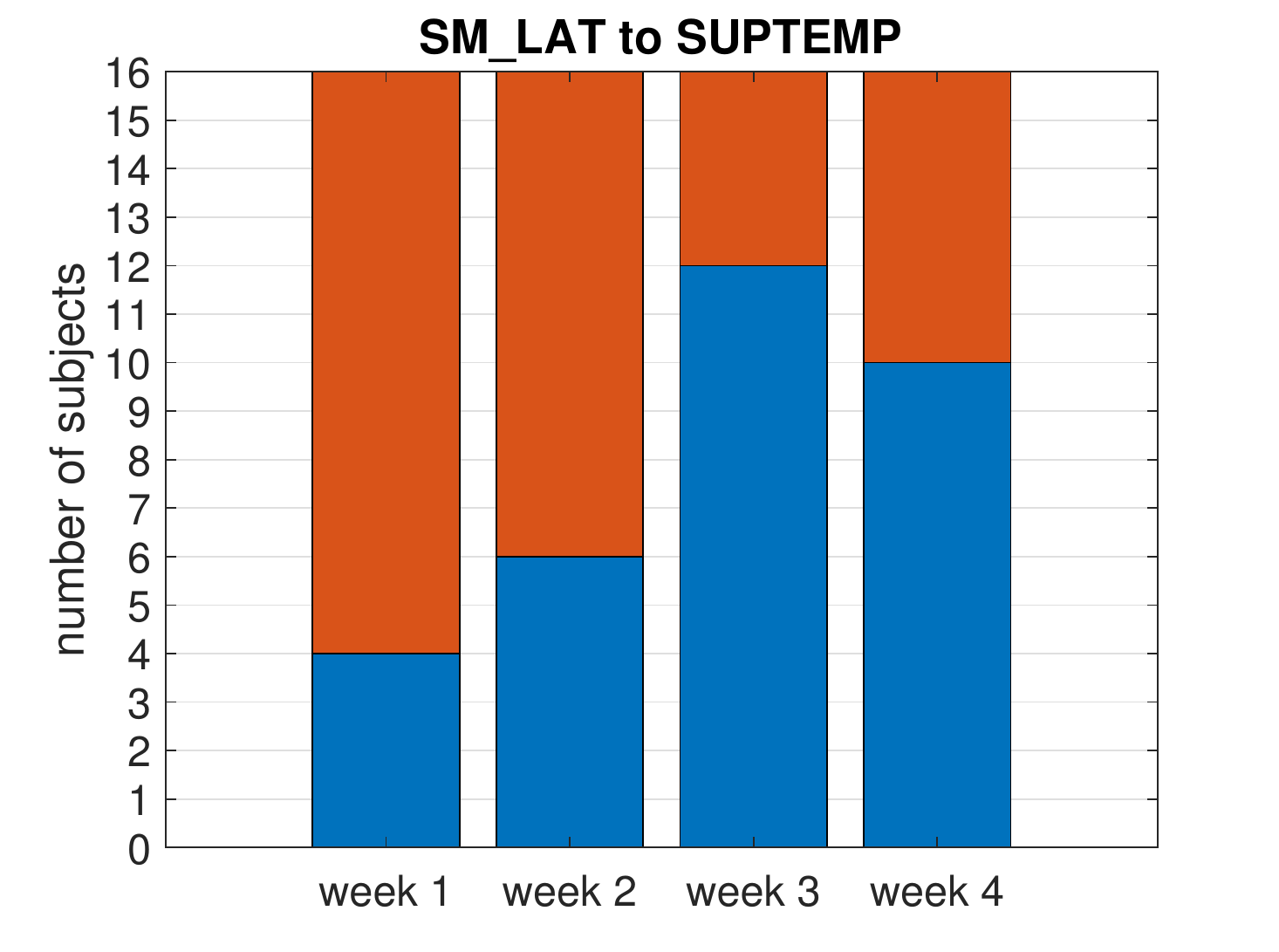}
    \caption{Connection from the lateral sensori-motor network to the superior temporal gyrus.}
    \label{fig:selfr3}
    \end{subfigure}
    \hfill
    \begin{subfigure}[t]{0.32\textwidth}
    \centering
    \includegraphics[width=\textwidth]{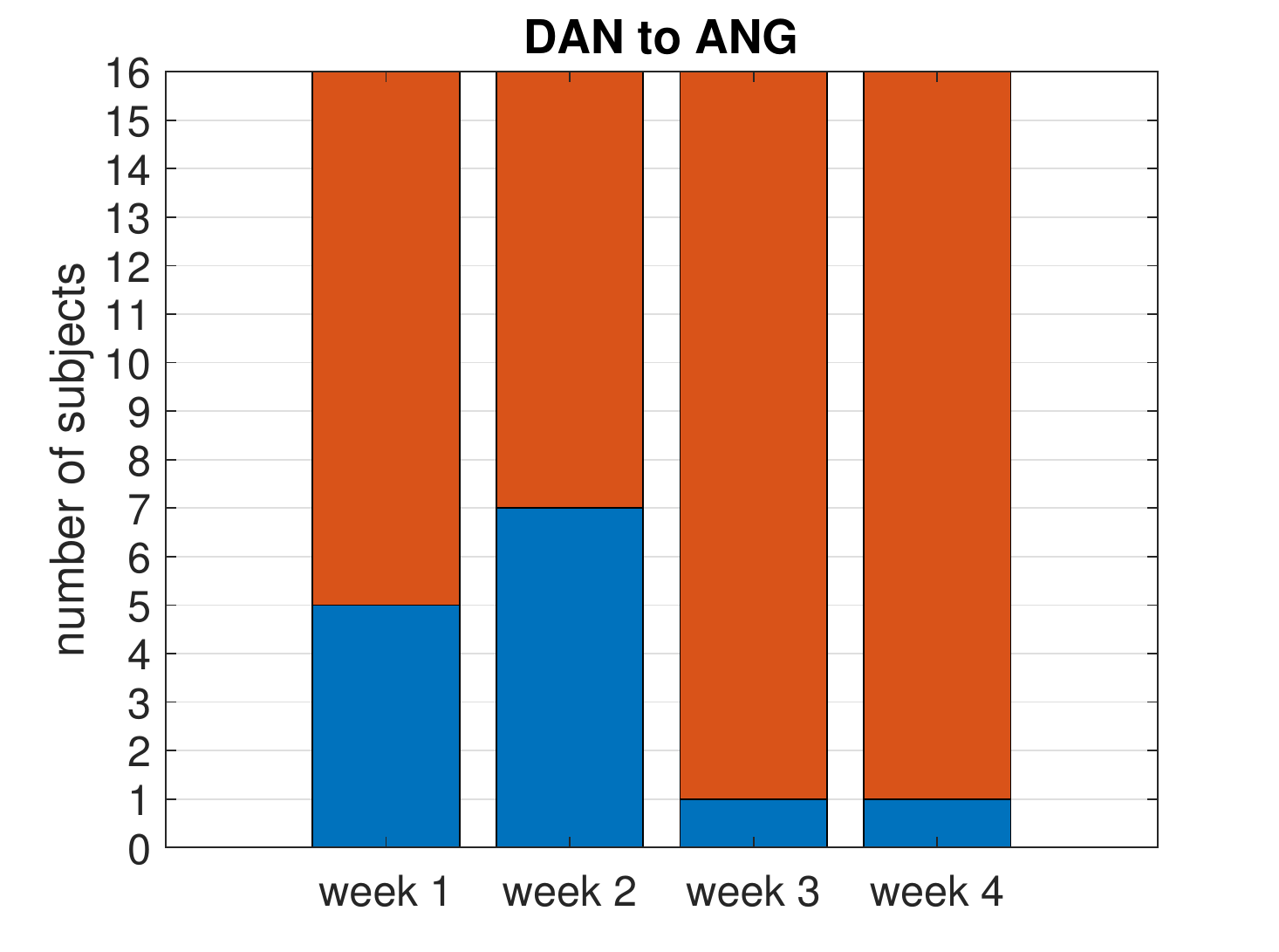}
    \caption{Connection from the dorsal attention network to the angular gyrus.}
    \label{fig:selfr4}
    \end{subfigure}
    \hfill
    \begin{subfigure}[t]{0.32\textwidth}
    \centering
    \includegraphics[width=\textwidth]{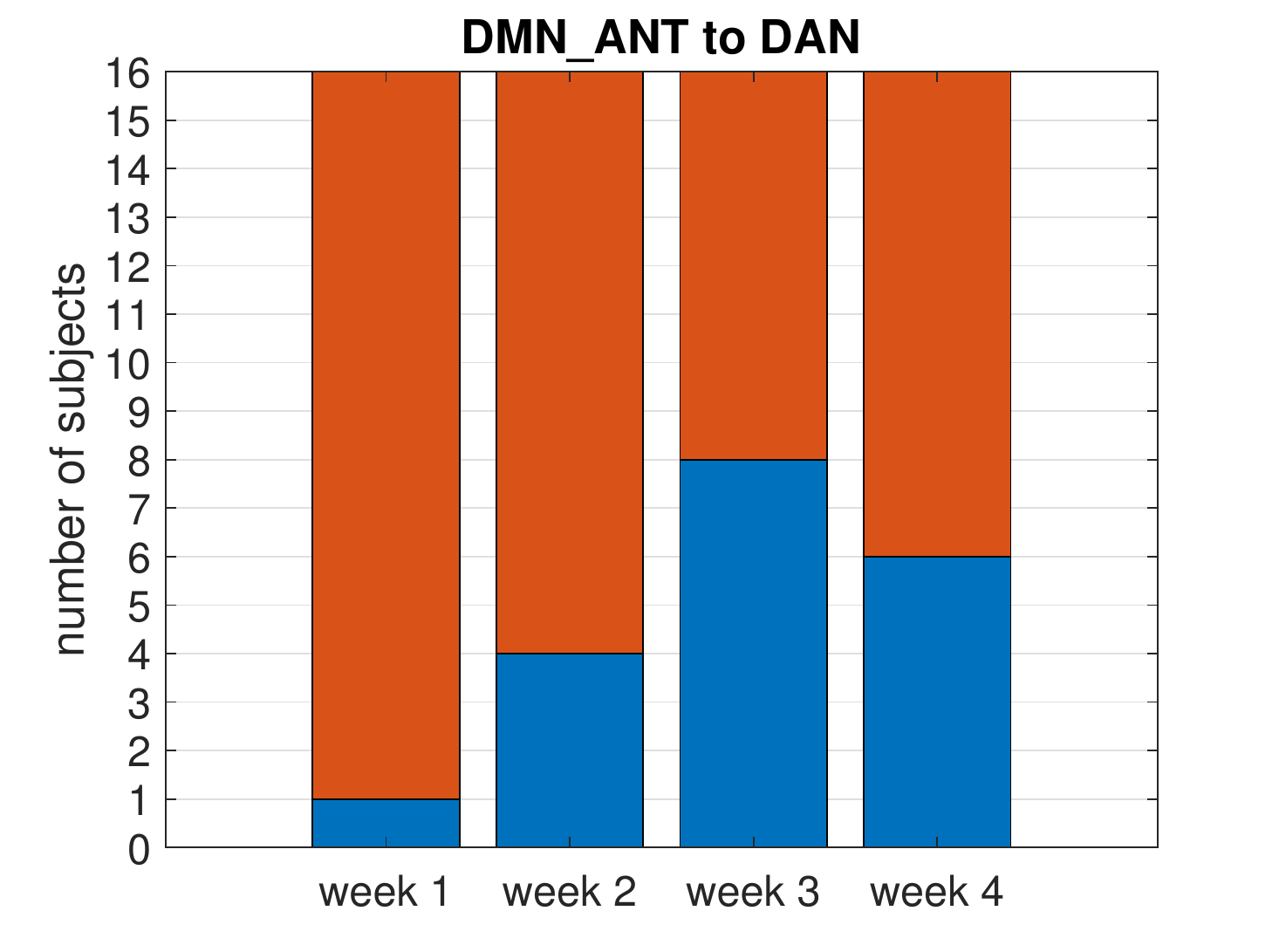}
    \caption{Connection from the anterior default mode network to the dorsal attention network.}
    \label{fig:selfr6}
    \end{subfigure}
    \caption{Selection frequencies of four connections of with possible effects of interest. Here, red indicates the number of subjects that did not have the connection in that week and blue indicates the number of subjects that did have the connection. (This is a 2-column figure)}
    \label{fig:selfr}
\end{figure*}
\subsection{Inference of the Mozart effect}
From the initial search for the Mozart effect using the Bayesian selection frequency of the connections over the four weeks, we have chosen six connections, in which the selection frequency indicate that there might be a change in the optimal hypothesis between week 2 and 3, i.e. before and after the subjects listened to Mozart music. The Bayesian selection frequencies of these connections are reported in Figure \ref{fig:selfr}.\par
In Figure \ref{fig:selfr1} and \ref{fig:selfr2} we found two connections inbound to the fronto-parietal right network and in \ref{fig:selfr6} we see the connection from the anterior default mode network to the dorsal attention network. These connections are of interest, because the default mode networks and the fronto-parietal right network are involved in cognitive processing \citep{MarekFPR,DMNRaichle}. In Figure \ref{fig:selfr5} we see the selection frequency of a connection involving the central executive network, which is involved in cognitive processing \citep{CENMiller}. Then, in Figure \ref{fig:selfr3} we show the selection frequency of the connection from the sensori-motor, lateral and the superior temporal gyrus. Figure \ref{fig:selfr3} is of special interest because first of all it has the largest selection frequency for any connection over all scans, with 12 out of 16 subjects showing positive evidence of the connection. Furthermore, it has a relatively small drop in selection frequency in week 4 relative to week 3. Both of the brain networks involved in the connection are involved in auditory processing in the brain \citep{SMlatRecruit,MusicFMRI}. Finally, in Figure \ref{fig:selfr4} we show a connection from the dorsal attention network to the angular gyrus. This connection is of interest, because a low variability large selection frequency was detected in week 1 and 2, which almost completely disappears in week 3 and does not increase at all in week 4. This could indicate that effective connectivity in connections can decrease as a result of listening to Mozart music. In summary, we have found a small selection of potential effects in our analysis of the selection frequency. However, while we see some changes in selection frequency between week 2 and 3, the effects are far from universal across all subjects.\par 
\begin{table*}[width=.9\textwidth,cols=4,pos=t]
    \centering
    \begin{tabular*}{\tblwidth}{@{} LLLL|LLL|LLL|LLL@{} }
    \toprule
    \multirow{2}{*}{Connections} & \multicolumn{3}{c|}{week 1} & \multicolumn{3}{c|}{week 2} & \multicolumn{3}{c|}{week 3} & \multicolumn{3}{c}{week 4} \\
     & $H_{opt}$ & BF & pos & $H_{opt}$ & BF & pos & $H_{opt}$ & BF & pos & $H_{opt}$ & BF & pos \\
    \midrule
    DMN\_ANT to DAN & $H_0$ & -40.08 & 1 & $H_0$ & -33.62 & 1 & $H_0$ & -7.305 & 0.97 & $H_0$ &  -29.65 & 1 \\
    DMN\_POS to FPR & $H_0$ & -41.26 & 1 & $H_0$ & -44.80 & 1 & $H_1$ & 1.890 & 0.72 & $H_0$ & -45.07 & 1 \\
    CEN to SM\_SUP & $H_0$ & -38.88 & 1 & $H_0$ & -47.75 & 1 & $H_1$ & 8.266 & 0.98 &  $H_0$ & -27.03 & 1 \\
    SM\_LAT to SUPTEMP & $H_0$ & -20.78 & 1 & $H_1$ & 37.46 & 1 & $H_1$ & 237.7 & 1 & $H_1$ & 72.34 & 1 \\
    DAN to ANG & $H_0$ & -24.73 & 1 & $H_0$ & -39.28 & 1 & $H_0$ & -109.5 & 1 & $H_0$ & -74.87 & 1 \\
    SM\_SUP to FPR & $H_0$ & -53.55 & 1 & $H_0$ & -70.83 & 1 & $H_0$ & -53.21 & 1 & $H_0$ & -86.71 & 1 \\
    \bottomrule
\end{tabular*}
    \caption{Results of the extended Bayesian method on the ICA time series in the Mozart effect study. The results are summarized for each of the four weeks, where we indicate for each week the optimal hypothesis, the Bayes factor (BF) and the posterior probability (pos) of the optimal hypothesis in that week.}
    \label{tab:hyp}
\end{table*}
Next, we assessed the optimal group hypothesis of the 16 subjects for each week for the six connection we reported in Figure \ref{fig:selfr}. We have presented the results in Table \ref{tab:hyp}. In Table \ref{tab:hyp} we report the optimal hypothesis of each connection for each week, its Bayes Factor BF and its posterior probability $p(H_{opt}|\mathcal{D})$. Now, from Table \ref{tab:hyp} we notice the connection from the posterior default mode network to the fronto-parietal right network and the connection from the central executive network to the sensori-motor superior network. For these two connections the optimal hypothesis in week 1 and 2 was $H_0$, and the optimal hypothesis in week 3 was $H_1$, indicating a change in effective connectivity of these connections after the subjects listened to Mozart music. Furthermore, in both connections the change did not last into week 4, as the optimal hypothesis was once again $H_0$. Next, we note in Table \ref{tab:hyp} that the connection from the sensori motor lateral network to the superior temporal gyrus shows strong evidence in favor of $H_1$ in week 3, however, the optimal hypothesis in week 1 and 2 changes from $H_0$ to $H_1$ and therefore we conclude that the natural variability of this connection before listening to Mozart music is too large to detect a change in effective connectivity caused by the Mozart music. Finally, we note that for the other three connections, the optimal hypothesis for both week 1 and 2 did not change, but there was also no change in optimal hypothesis between week 2 to week 3 and as such there was no significant change in the effective connectivity of the connections in the group of all 16 subjects. The brain networks involved in the connections for which a change in optimal hypothesis occurred only between week 2 and 3, have been summarized in Figure \ref{fig:networks_vis}. \par
Finally, for all six connections in Figure \ref{fig:selfr}, we found no significant difference in Granger-values between week 1 and 2. Then we compared the Granger-values of week 2 and 3 and found significance of the paired t-test only for the connection from the anterior default mode network to the dorsal attention network with $p=0.035$. Next, we inferred only for this connection if there are any differences in Granger-values between week 3 and 4 and found no significant difference between the two weeks, indicating that the effective connectivity change persisted into week 4.\par
\begin{table*}[width=.9\textwidth,cols=4,pos=t]
    \centering
    \begin{tabular*}{\tblwidth}{@{} LLLL|LLL|LLL|LLL@{} }
    \toprule
    \multirow{2}{*}{Connections} & \multicolumn{3}{c|}{week 1} & \multicolumn{3}{c|}{week 2} & \multicolumn{3}{c|}{week 3} & \multicolumn{3}{c}{week 4} \\
    %\cline{2-13}
     & $H_{opt}$ & BF & pos & $H_{opt}$ & BF & pos & $H_{opt}$ & BF & pos & $H_{opt}$ & BF & pos \\
     \midrule
    DMN\_ANT to DAN & $H_0$ & -15.09 & 1 & $H_0$ & -25.46 & 1 & $H_1$ & 2.43 & 0.77 & $H_0$ & -28.49 & 1\\
    DMN\_POS to FPR & $H_0$ & -21.37 & 1 & $H_0$ & -56.62 & 1 & $H_1$ & 13.77 & 0.99 & $H_0$ & -10.76 & 1\\
    \bottomrule
\end{tabular*}
    \caption{Results of the inference based on the listening duration of subjects from the Bayesian method. The results are summarized for each of the four weeks, where we indicate for each week the optimal hypothesis, the Bayes factor (BF) and the posterior probability (pos) of the optimal hypothesis in that week. Only two connections are shown here, which both only showed a change in optimal hypothesis only for the subgroup of longer listening subjects. The evidence and optimal hypothesis of the shorter listening subgroup is not shown as they contain no effects of interest.}
    \label{tab:hyplisdur}
\end{table*}
\subsubsection{Inference based on listening duration}
In our final test, the subjects are divided into two groups of 8 subjects. The subjects in group 1 listened on average 27:14$\pm$7:07 hours to Mozart music between week 2 and 3 and the subjects in group 2 listened on average 16:19$\pm$1:31 hours to Mozart music. Both methods were repeated for the new subgroups.\par
Now, using the Bayesian hypothesis test on both subgroups, we have found two connections that showed an effect only for the subgroup with a longer listening duration, but no effect for the group with a shorter listening duration. These two connections are presented in Table \ref{tab:hyplisdur}. We only show the BF and optimal hypotheses for the subgroup which listened for a longer duration, as no change in optimal hypothesis was found between week 2 and 3 for the shorter listening subgroup. The other connections which were selected in Figure \ref{fig:selfr}, were not favored by one of the subgroups over the other and are therefore also not presented here.\par
\begin{figure*}
\centering
\begin{framed}
    \begin{subfigure}[t]{0.32\textwidth}
    \centering
    \caption*{DMN\_POS to FPR}
    \includegraphics[width=\textwidth]{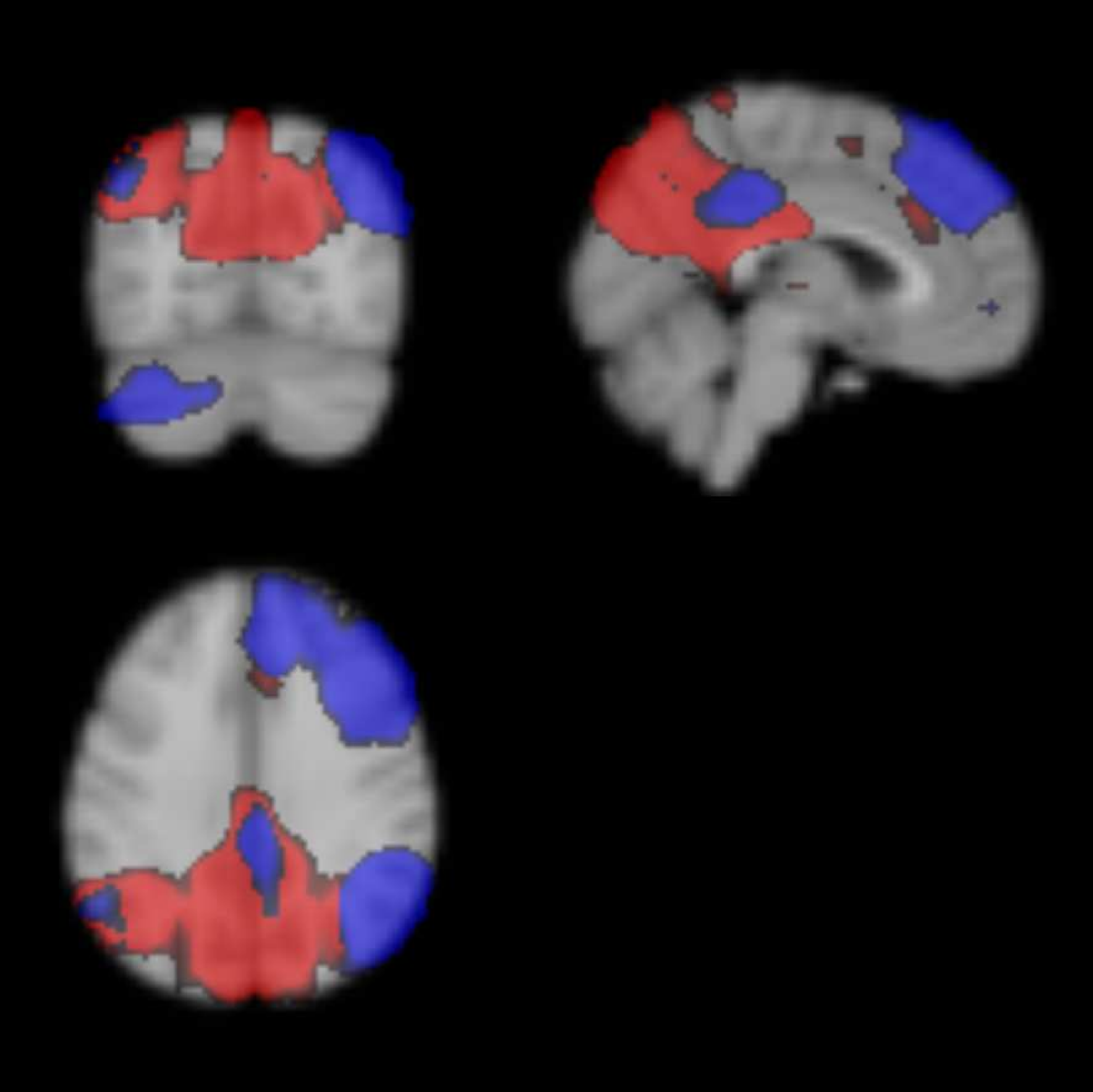}
    \end{subfigure}
    %\hfill
    \begin{subfigure}[t]{0.32\textwidth}
    \centering
    \caption*{DMN\_ANT to DAN}
    \includegraphics[width=\textwidth]{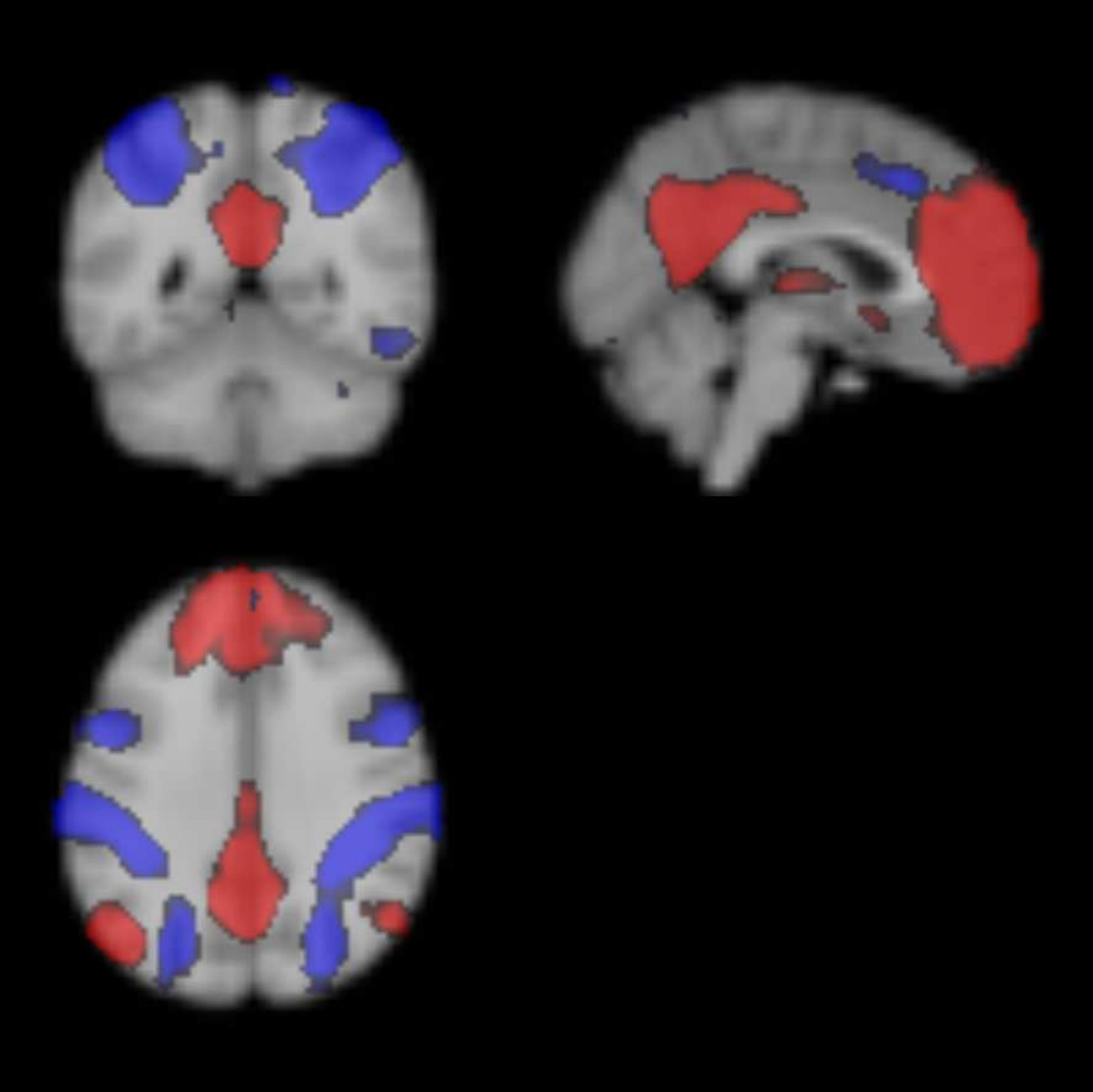}
    \end{subfigure}
    %\hfill
    \begin{subfigure}[t]{0.32\textwidth}
    \centering
    \caption*{CEN to SM\_SUP}
    \includegraphics[width=\textwidth]{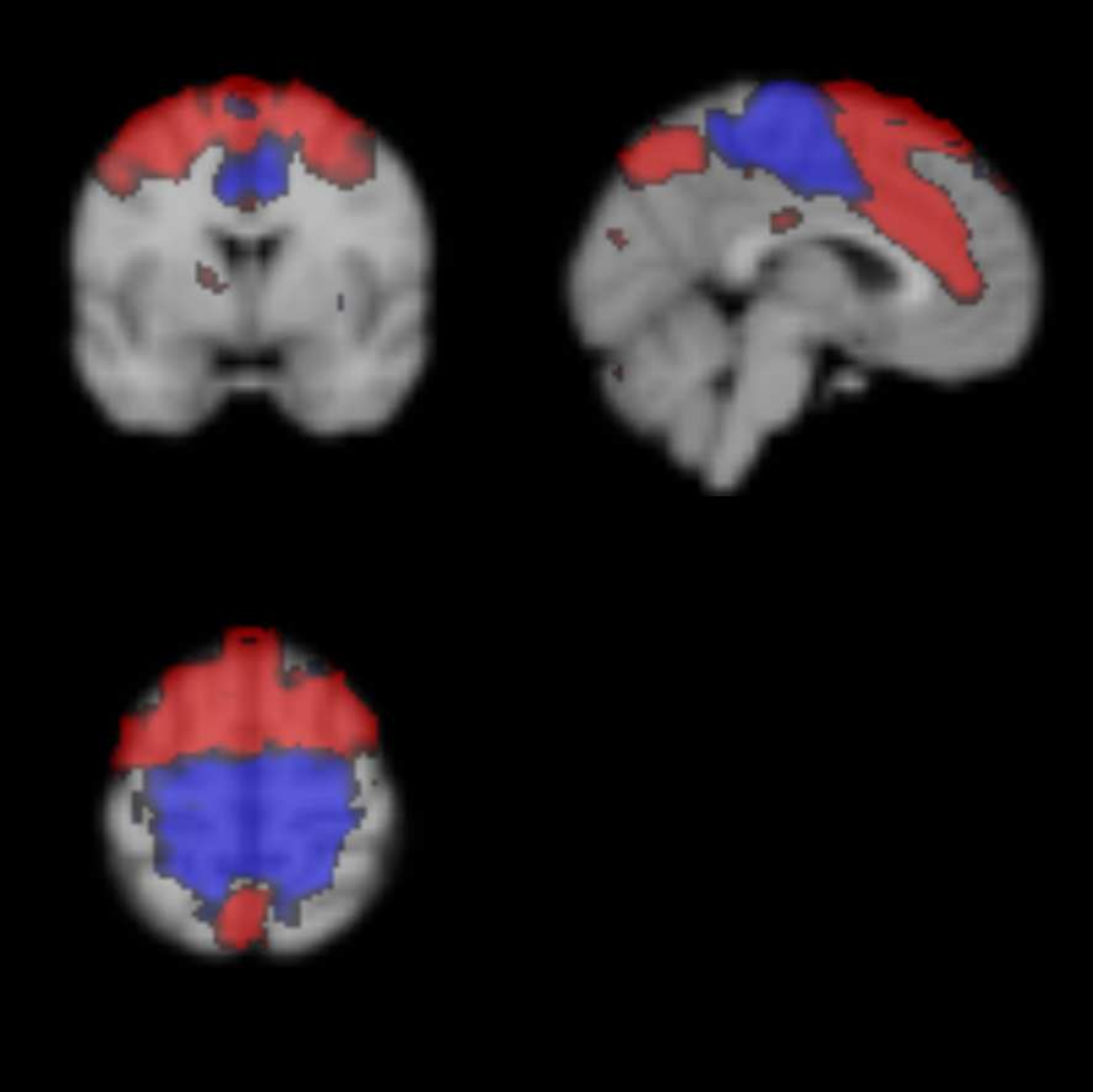}
    \end{subfigure}
    %\hfill
\end{framed}
    \caption{Brain networks in which the connections which were found to have the strongest evidence of a change in effective connectivity after subjects (full group or long listening subgroup) listened to Mozart music (see row 1-3 in Table \ref{tab:hyp} and in Table \ref{tab:hyplisdur}; Figure \ref{fig:selfr1}, \ref{fig:selfr5} and \ref{fig:selfr6}). Here, the direction of the connection is from the red network to the blue network. (This is a 2-column figure)}
    \label{fig:networks_vis}
\end{figure*}

The first connection in Table \ref{tab:hyplisdur}, from the anterior default mode network to the dorsal attention network, which showed no change in optimal hypothesis between week 2 and 3 for the group of 16 subjects, shows positive evidence in favor of $H_1$ in week 3 for the subgroup of longer listeners. As such there is a change in optimal hypothesis between week 2 and 3 for the longer listening subgroup. The change in optimal hypothesis between week 2 and 3 did not last in week 4, as such this effect was of short duration. The second connection in Table \ref{tab:hyplisdur}, from the posterior default mode network to the fronto-parietal right showed weak evidence in favor of $H_1$ in week 3 for the group of 16 subjects and as such there was a change in optimal hypothesis between week 2 and 3. This connection shows strong evidence in favor of $H_1$ in week 3 for the subgroup of longer listeners, as such the change in optimal hypothesis between week 2 and 3 in the full group of 16 subjects was predominantly caused by the longer listening subjects. Similarly to the inference in the full group of 16 subjects, the change in optimal hypothesis between week 2 and 3 did not last into week 4, thus the effect was of short duration. A visualisation of the networks involved in these connections can be found in Figure \ref{fig:networks_vis}. \par 
Finally, no results were found for the significance of Granger-values for both groups. All comparisons between weeks for both long and short listeners had no significant change in effective connectivity between any of the weeks for all selected connections in Figure \ref{fig:selfr}.\par 
\section{Discussion}
In this study, we illustrated, using simulation data, that the Bayesian topology identification outperforms the Granger-causality analysis, especially when not much data is available. The Bayesian method was extended to enable it to test hypotheses on the existence of a connection for each week separately. The optimal hypotheses of each week are combined to draw conclusions on changes in the effective connectivity over the weeks in the Mozart study. Then we applied the extended Bayesian method and the Granger-causality analysis to the Mozart effect data and found a number of connections with changes in effective connectivity after the subject listened to Mozart music. Finally, we split the subjects into two groups based on listening duration and found that for two connections a change in effective connectivity could only be detected in the subgroup of longer listening subjects.\par 
\subsection{Simulations}
We evaluated the Bayesian topology identification as an alternative to the popular Granger-causality analysis for the inference of brain network connectivity. The methods are compared through their performance in estimating a graph of the effective connectivity using simulated data with a known ground truth connectivity. Herein, we expected that the Bayesian method should have an advantage in the inference over the Granger-causality analysis through the use of a parameter prior and by modelling the AR model parameters as random variables. The performance measurements in Figure \ref{fig:simperf} show that the Bayesian method performs better than the Granger-causality analysis for short data lengths using the simulation data. As the data length increased we saw that the Granger-causality analysis performance more closely matched that of the Bayesian method. Thus, for fMRI data studies, where often not many data points are available due to the slow sampling rate of fMRI measurements, it seems that the Bayesian method has a distinct advantage in the inference of brain network connectivity over Granger-causality analysis. \par
\subsection{Mozart effect study}
In the study of the Mozart effect, we applied both the Bayesian method and the Granger-causality analysis to the ICA time series of the subjects. Here, we focused on the Bayesian method, because we observed it performed better than the Granger-causality analysis in the evaluations of performance using simulated data. First, using the Bayesian hypothesis test we detected a change in effective connectivity of multiple connections after subjects listened to Mozart music. For the full group of 16 subjects, we detected a change in optimal hypothesis for the connection from the central executive network to the dorsal attention network, which was of short duration. The central executive network is involved in maintaining and manipulating working memory and performs goal-oriented decision making \citep{CENMiller}. Therefore the change in effective connectivity could be an indication of a change in cognition.\par
Next, the connection from the posterior default mode network to the fronto-parietal right network also showed an increase in effective connectivity after listening to music as measured by the Bayesian method, although the evidence that $H_1$ was the optimal hypothesis in week 3 was only weak. When we split the subjects into subgroups based on listening duration, it becomes apparent that the effect predominantly occurs in the longer listening subjects and was of short duration. The default mode network is involved in emotional processing, self-referential mental activity, and the posterior default mode network in particular involves recollection of past experiences \citep{DMNRaichle}. The fronto-parietal right network is involved in cognitive control \citep{MarekFPR}. This change in effective connectivity between two networks involved in cognitive processing is an interesting result as it could indicate the existence of the Mozart effect.\par
Finally, the connection from the anterior default mode network to the dorsal attention network was detected to have a significant change in effective connectivity between week 2 and 3 by the Granger-causality analysis for the whole group of subjects. In the Bayesian method a change in optimal hypothesis was detected, but only for the subgroup of longer listening subjects. However, the two methods disagree on whether the effect was of short duration or not and as such we cannot draw any strong conclusions on the duration of the change. Overall, the evidence that this connection was present in week 3 from the Bayesian method was not strong and the connection was only present in the subgroup. However, it is supported by the Granger-causality analysis results, and as such we are more confident in the result that an actual change in effective connectivity happened after listening to Mozart music. We have mentioned the functionality of the default mode network already above. Furthermore, the dorsal attention network is involved in the orientation of attention to external stimuli \citep{DANVossel}. Thus the brain networks in this connection are involved in cognitive processing. The change in effective connectivity caused by Mozart music, detected by both methods, could indicate evidence of the Mozart effect.
\par
\subsection{Limitations and future work}
First, compared to the Granger-values, which are a measurement of connectivity strength and for which statistical tests exist to infer a difference in connectivity between groups of subjects, the extended Bayesian method can only test binary hypotheses of the effective connectivity. 
In future development of the Bayesian method, one of our goals will be to extend the Bayesian hypothesis test to provide a quantitative measure of the connectivity strength. This could help us draw conclusion on effective connectivity change in the connection from the sensori-motor lateral network to the superior temporal gyrus, where even though the optimal hypothesis was $H_1$ in both week 2 and 3, we see a very large increase in the evidence BF in week 3 as compared to week 2, but with our currently available binary hypotheses on effective connectivity, we cannot confirm if this indicates a change or not. \par
%
%Mozart effects were detected after 2-week listening time in \citet{MozartCoppola}, and thus the listening duration of this work was also designed in the scale of weeks. We considered this study to be explorative and thus chose $1$-week listening duration for the subjects.
%
Second, although we had originally mandated a minimum of 2 hours a day of listening to music, this minimum listening duration appeared to be insufficient to cause changes in effective connectivity in all subjects for the connections involving the anterior and posterior default mode networks. The listening duration in other studies of the effect of Mozart music on epileptic seizure frequency was at least 2 hours a day for 15 days \citep{MozartCoppola}. In another study \citep{MozartBodner} patients were exposed to Mozart music for a year. 
Therefore, in future work we would like to increase the minimum listening duration of subjects, to verify whether in that case effects will be more universal across subjects. Furthermore, because listening to Mozart music for more than 2 hours a day might be hard to accommodate for subjects, it might be necessary to increase the time interval between scan 2 and 3 to reduce the daily listening time to a manageable duration.\par
Finally, we considered this study to be exploratory, and potential future work can consider an increased number of subjects, now that the insight of this work has been obtained.\par 
\section{Conclusions}
This was an exploratory study of the Mozart effect by inferring changes in brain network connectivity using ICA time series from fMRI data. As far as we are aware, this study is the first of its kind in measuring effective connectivity changes caused by Mozart's Sonata K448. Furthermore, it is the first time that the Bayesian topology identification algorithm is applied to fMRI data. We have found changes in cognitive processing in subjects, some of which were predominantly in the subjects with a longer listening duration. More effects were hinted at by the Bayesian selection frequencies in Figure \ref{fig:selfr}, but could ultimately not be detected through the Bayesian hypothesis test. We are hopeful that in future studies, with increased listening duration and more subjects, more changes in effective connectivity caused by Mozart music will be found.
\section*{Declaration of competing interests}
None.
\section*{Acknowledgements}
This project has received funding from the European Research Council (ERC), Advanced Research Grant SYSDYNET, under the European Union’s Horizon 2020 research and innovation programme (grant agreement No 694504). Shengling Shi would like to thank Maarten Schoukens for his input into this work.\par
\appendix
\section{Simplification of likelihood marginalization}
\label{appendA}
First of all, note that from the definition of our model in \eqref{eq:ARvecform} and $p(D|\mathcal{G})$ in \eqref{eq:margll}, we see that the marginalization in \eqref{eq:graphmargll} can be decomposed into $L$ separate marginalizations. Then, we only need to marginalize over $\mathcal{G}_j$, i.e. the subgraph which contains connection $e_{ji}$:
\begin{equation}\label{eq:graphmargMISO}
    p(D^k|e_{ji})\propto
    \sum_{\mathcal{G}_j\in \mathcal{P}_{j,1}}
    p(\mathcal{G}_j)p(D^k|\mathcal{G}_j),
\end{equation}
where $\mathcal{P}_{j,1}=\{\mathcal{G}_j|e_{ji}\in \mathcal{G}_j\}$. Note that $p(D^k|e_{ji})$ is now only proportional to the marginalization. This is not an issue, because when we use the proportional marginals from \eqref{eq:graphmargMISO} to calculate $\operatorname{BF}$ in \eqref{eq:BF}, it will still result in the correct log-likelihood ratio.
\bibliography{reference}
\end{document}